\documentclass[a4paper,fleqn,usenatbib,useAMS]{mnras}

\usepackage{graphicx}	
\usepackage{amsmath}	
\usepackage{amssymb}	
\usepackage{multicol}        
\usepackage{longtable}

\usepackage{bm}		
\usepackage{pdflscape}	

\usepackage{enumerate}
\usepackage[shortlabels]{enumitem}
\usepackage{soul}

\defcitealias{McClure-Griffiths_2010}{McG10}

\usepackage[T1]{fontenc}
\usepackage{ae,aecompl}

\usepackage{newtxtext,newtxmath}

\title[RM towards the Magellanic Leading arm]{Distant probes of RM structure - Where is the Faraday Rotation towards the Magellanic Leading Arm?}

\author[Jung et al.]{S. Lyla Jung$^{1}$\thanks{e-mail: \href{mailto:lyla.jung@anu.edu.au}{lyla.jung@anu.edu.au}}, N. M. McClure-Griffiths$^{1}$, Alex S. Hill$^{2}$
\\
\\
$^{1}$ Research School of Astronomy \& Astrophysics, The Australian National University, Canberra ACT 2611, Australia
\\
$^{2}$ Department of Computer Science, Math, Physics, and Statistics, the University of British Columbia, Okanagan Campus, 3187 University Way, Kelowna,\\ BC V1V1V7, Canada}

\date{Last updated}

\pubyear{2021}

\begin{document}
\label{firstpage}
\pagerange{\pageref{firstpage}--\pageref{lastpage}}
\maketitle

\begin{abstract}
Faraday Rotation Measures (RM) should be interpreted with caution because there could be multiple magneto-ionized medium components that contribute to the net Faraday rotation along sight-lines.
We introduce a simple test using Galactic diffuse polarised emission that evaluates whether structures evident in RM observations are associated with distant circumgalactic medium (CGM) or foreground interstellar medium (ISM).
We focus on the Magellanic Leading Arm region where a clear excess of RM was previously reported. There are two gaseous objects standing out in this direction: the distant Magellanic Leading Arm and the nearby Antlia supernova remnant (SNR). We recognized narrow depolarised filaments in the $2.3\,\rm GHz$ S-band Polarization All Sky Survey (S-PASS) image that overlaps with the reported RM excess.
We suggest that there is a steep gradient in Faraday rotation in a foreground screen arising from the Antlia SNR.
The estimated strength of the line-of-sight component of the magnetic field is $B_{\parallel}\sim 5\,\rm\mu G$, assuming that the excess of RM is entirely an outcome of the magnetized supernova shell. Our analysis indicates that the overlap between the RM excess and the Magellanic Leading Arm is only a remarkable coincidence.
We suggest for future RM grid studies that checking Galactic diffuse polarisation maps is a convenient way to identify local Faraday screens.
\end{abstract}





\section{Introduction}

Faraday rotation enables studies on the cosmic magnetism.
As linearly polarised radiation travels along a line-of-sight, its plane of polarisation rotates through an angle $RM \lambda^{2}$, where
\begin{equation}\label{eq:RM}
    RM = 0.812\int_{\rm observer}^{\rm source}n_{\rm e}(r)B_{\parallel}(r)dr,
\end{equation}
where RM is in units of $\rm rad\,m^{-2}$, $n_{\rm e}$ is electron density in $\rm cm^{-3}$, $B_{\parallel}$ is the magnetic field strength along the line-of-sight in $\mu G$, and r is a path length in $\rm pc$.
Extragalactic compact sources, e.g., radio galaxies and quasars, are often used to probe RM at a pinpoint location on the sky plane.
The integration is between the source and the observer along the line-of-sight; any patches of magneto-ionic media that polarised radiation is transmitted through on the way to the observer cause Faraday rotation and influence the resulting RM.

In practice, it is common to have multiple Faraday rotating regions along a line of sight within a telescope beam. The RM of such 'Faraday complex' sources is expressed using the Faraday depth ($\phi$), a parameter that describes the Faraday rotation at individual Faraday screens (\citealt{Burn_1966});
\begin{equation}\label{eq:faraday_rotation}
    \phi(X) = 0.812\int_{\rm observer}^{\rm X}n_{\rm e}(r)B_{\parallel}(r)dr,
\end{equation}
where $X$ is a certain position along the line-of-sight and $\phi$ is a function of $X$. In this study, we use RM and $\phi$ interchangeably. Details will be described in Section \ref{sec:atca}.

The `RM grid' technique probes a contrast in the overall distribution of RM on and off magneto-ionic structures, assuming RM sources are mostly in the background of target objects. For this statistical approach, a sufficient number of polarised sources is required in the region of interest.
Thus far, the RM source density of current radio surveys (e.g., $1 \,\rm deg^{-2}$, \citealt{Taylor_2009}) limits this approach on studies of the large-scale Galactic magnetic fields or several square-degree-size extended objects.
Upcoming radio surveys using next generation radio telescopes are expected to provide immensely denser RM grid (e.g., $25\,\rm deg^{-2}$, \citealt{Anderson_2021}). 
Ever-improving RM grids are enabling new measurements of the magnetic field strength and structure in a range of objects including galaxy clusters (e.g. \citealt{Anderson_2021}), individual resolved galaxies (e.g. \citealt{Gaensler_2005}; \citealt{Mao_2008}), Galactic objects (e.g. \citealt{Harvey-Smith_2011}) and high velocity clouds (e.g. \citealt{McClure-Griffiths_2010}; \citealt{Hill_2013}).


There is ample evidence that we are surrounded by the multi-phase circumgalactic medium (CGM) in the Milky Way halo (see \citealt{Putman_2012} and references therein).
Understanding the nature of the CGM is key to understanding the evolution of galaxies, as it connects the pristine intergalactic medium (IGM) to the star-forming interstellar medium (ISM).
The Milky Way provides a unique environment for studying the complex structures of the CGM on a spatially resolved scale, which is hard to achieve by studies of distant galaxies.
Although most baryons belonging to the Milky Way are distributed in the Galactic disc, observations have reported the presence of extraplanar gas clouds at high Galactic altitude moving at a high relative velocity with respect to Galactic rotation (\citealt{Muller_1963}; \citealt{Putman_2002}; \citealt{McClure-Griffiths_2009}; \citealt{Saul_2012}). They are so-called High-Velocity Clouds (HVCs) and Intermediate-Velocity Clouds (IVCs).

As HVCs travel through the halo, they hydrodynamically interact with their surroundings. Ram pressure pushes and strips low-density structures to the opposite direction of the cloud motion, Kelvin-Helmholtz and Rayleigh-Taylor instabilities 
develop turbulent mixing layer across the cloud-wind interface, which is susceptible to the cloud stripping (\citealt{Jones_1994}; \citealt{Schiano_1995}). 
Taken all together, it is easy to conclude that moving clouds dissipate in a short timescale and if so, one would expect to find HVCs dominated in the warm ionized phase (\citealt{Heitsch_2009}).
However, a significant mass of observed HVCs are in cold phase HI and they are often suggested as a source of cold gas fueling the star-formation of the Milky Way (\citealt{Putman_2012}).
This indicates additional forces that stabilize HVCs so that the clouds survive their journey through the Galactic halo and deliver gas to the Galactic disc.

The magnetic field is proposed as one of the possible sources that provide stability of HVCs (\citealt{Konz_2002}; \citealt{Santillan_2004}; \citealt{Kwak_2011}; \citealt{McCourt_2015}; \citealt{Banda-Barragan_2016};\citealt{Banda-Barragan_2018}; \citealt{Gronnow_2017}; \citealt{Gronnow_2018}).
Magnetic fields can be dragged along and amplified following the motion of plasma they are embedded in.
This dragging effect leads to the field `draping' around the moving clouds when the relative velocity between the ambient medium and the clouds is high enough to overcome the tension of the ambient magnetic field (\citealt{Dursi_2008}).
Hence, the magnetic field draping is particularly efficient around HVCs due to the high relative velocity and the weak magnetic fields of the Galactic halo.

There have been attempts to observationally constrain the magnetic field strength associated with the Miky Way HVCs using the RM grid (\citealt[hereafter, McG10]{McClure-Griffiths_2010}; \citealt{Hill_2013}; \citealt{Kaczmarek_2017}; \citealt{Betti_2019}).
In this paper, we present a high-density RM grid towards the Magellanic Leading Arm/Antlia SNR regions. We combine it with the Galactic diffuse polarised emission in order to test 
whether the Magellanic HVC or the Antlia SNR is the source of Faraday rotation.
The paper is structured as follows: in Section \ref{sec:method}, we present an overview of observation data and the Faraday RM synthesis technique.
In Section \ref{sec:confusion}, we introduce properties of two gaseous objects-- the Antlia SNR and the Magellanic Leading Arm-- overlapping closely with each other.
Then we explore the compact-source RM (Section \ref{Sec:RM}) and the diffuse continuum emission (Section \ref{sec:diffuse}) in the region. 
The discussions on our results are presented in Section \ref{sec:discussion}, including Section \ref{sec:b_los} where we estimate the magnitude of the line-of-sight magnetic field of Antlia SNR based on the observation data presented. 
Our conclusions are summarized in Section \ref{sec:summary}.

\section{Methodology}\label{sec:method}

\subsection{Compact-source rotation measure}\label{sec:rm_cat}

For constructing the RM grid covering our field-of-interest, we use two publicly available RM catalogues in addition to new radio continuum observations with the Australia Telescope Compact Array (ATCA). A summary of observational specifications are presented in the following subsections.

\subsubsection{NVSS and S-PASS/ATCA RM catalogue}
First, we have a RM catalogue from the NRAO Very Large Array Sky Survey (NVSS, \citealt{Condon_1998}) RM catalogue (\citet{Taylor_2009}) that covers the sky above a declination of $-40^{\circ}$ with the RM source density of $1\,\rm deg^{-2}$ on average.
Although it has a significant role in the search for magnetised HVCs (\citetalias{McClure-Griffiths_2010}; \citealt{Hill_2013}), it should be noted that the RMs are derived from limited coverage in frequency domain: $42\,\rm MHz$-wide bands centred at $1364.9\,\rm MHz$ and $1435.1\,\rm MHz$. Yet, the  $n\phi$-ambiguity tests performed by \citet{Ma_2019} has shown that the NVSS RM catalog is mostly reliable for sources located out of the Galactic plane and in the $\phi$ range used for this study.

For the southern sky below a declination of $-1^{\circ}$, there is the S-band polarisation All-Sky Survey (S-PASS, \citealt{Carretti_2019}). S-PASS/ATCA (\citealt{Schnitzeler_2019}) catalogue is derived from a follow-up observations of sources selected from S-PASS using ATCA. It provides the first wide-band ($1.3-3.1\,\rm GHz$) polarimetry data of compact sources with the average polarised source number density of $0.2\,\rm deg^{-2}$. 

\subsubsection{ATCA observations}\label{sec:atca}

\begin{figure}
    \centering
    \includegraphics[width=0.9\columnwidth]{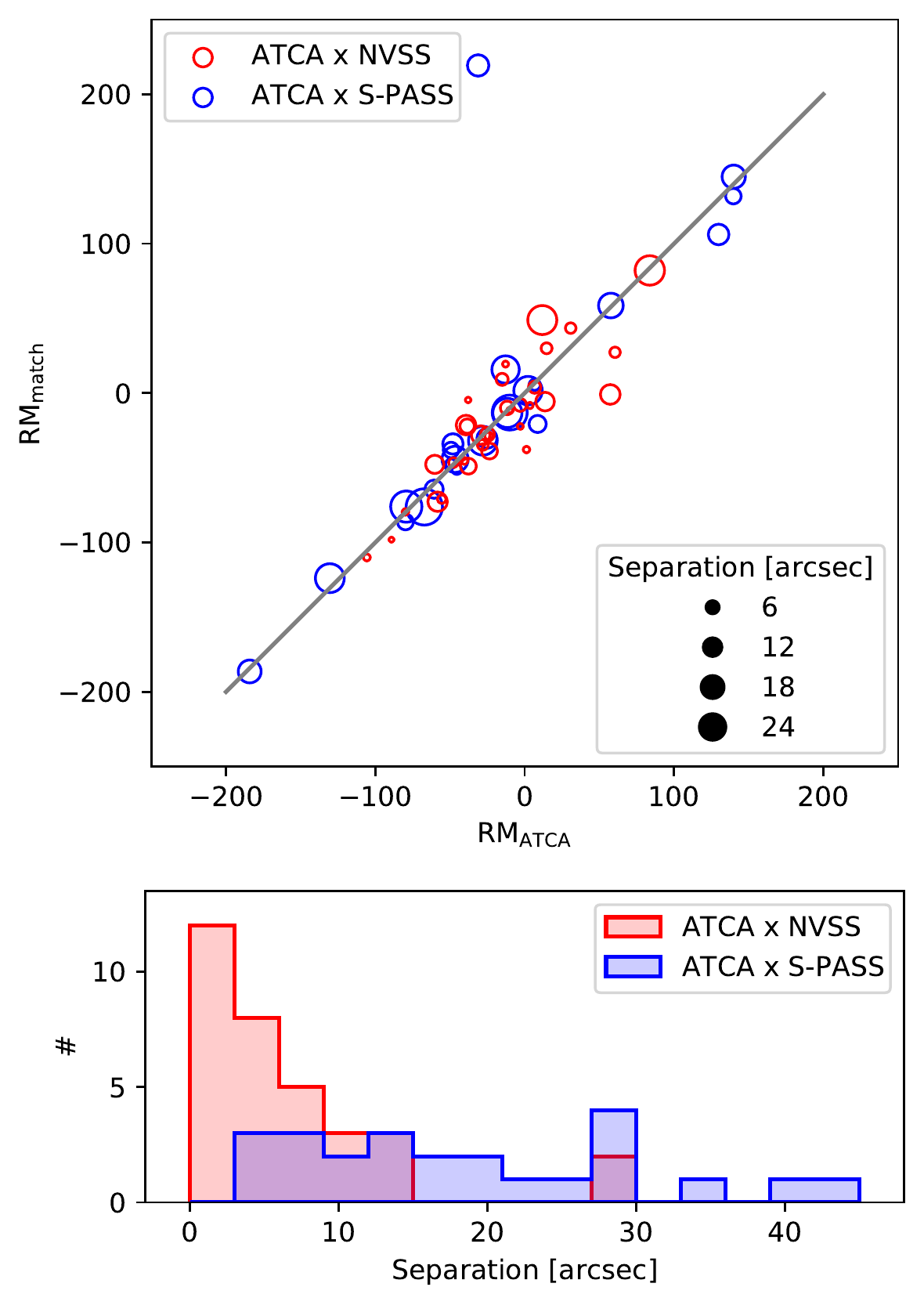}
    \caption{
    Top: Comparison of RMs matched between different catalogues using sky coordinates (red: ATCA $\times$ NVSS, blue: ATCA $\times$ S-PASS). The size of the circles represents the angular separation between matched pairs. The grey line shows the one-to-one relation where RMs from the two catalogues are identical.
    Bottom: The histogram of the separation between the matched pairs.
    }  
    \label{fig:rm_matching}
\end{figure}

In order to increase the polarised source density, we performed follow up observations on 737 fields in the region $10h:00m:00s < \alpha< 13h:30m:00s$ and $-52.3^{\circ} < \delta < -32^{\circ}$ with ATCA. Each source was visited $\approx6$ times over a 12-hour scan and the total observation time is on average 1.5 minutes per source.
The frequency range used for our analysis is $2\,\rm GHz$ wide continuum band from $1.1\,\rm GHz$ to $3.1\,\rm GHz$, and the spectral resolution is $\approx 1\,\rm MHz$. The angular diameter of the largest beam at the lowest frequency is $13.2''$. Radio bright sources were identified using {\sc Aegean} source finder (\citealt{Hancock_2012}; \citealt{Hancock_2018}) based on Stokes $I$ clean image stacked over the entire 2GHz-wide band. The total number of sources detected in the field is about 3000, including both polarised and unpolarised sources.

We used the {\sc miriad} software package (\citealt{Sault_1995}) provided by Australia Telescope National Facility (ATNF) for data reduction and imaging. The software is particularly designed for processing radio interferometry data observed with ATCA.
Prior to imaging, we performed flagging using the \texttt{mirflag} task in {\sc miriad} which immediately flags channels whose amplitudes deviate more than 14 times the median deviation from the channel median.
For calibration of data, we used \texttt{1934-638} as a flux calibrator and \texttt{1104-445}, \texttt{1206-399}, and \texttt{1215-457} as phase calibrators, respectively for each night. The calibration process determines tables of bandpass functions, antenna gains, and polarisation leakage and corrects the observed visibility to get the ideal sky intensity distribution of radio sources.

Imaging of data was performed separately for each polarisation (Stokes $I, Q, U,$ and $V$) and each chunk of channels, where the channel width ($\delta \lambda^{2}$ in the equation \ref{eq:phi_max} below) is determined by estimating the maximum value of observable Faraday depth:
\begin{equation}\label{eq:phi_max}
    \left\| \phi_{\rm max}\right\| \approx \sqrt{3}/\delta \lambda^{2}.
\end{equation}
The targeted sky is at relatively high galactic latitude where we do not normally expect to detect extremely high Faraday rotation compared to the Galactic disc as both the magnetic field strength and ionized gas density is relatively low. Therefore, we used a channel width of $20\,\rm MHz$ which limits the measurable Faraday depth to around $\left\| \phi_{\rm max} \right\| \approx 750\, \rm rad\,\rm m^{-2}$.
In order to match the spatial resolution throughout the channels, the clean images of the frequency slices were smoothed by the largest beam among them before being stacked into a cube. At the end of this imaging stage, we have Stokes $I, Q, U, V$ parameters of the identified sources as a function of frequency.

The RM of each source was estimated using the Faraday RM synthesis technique (e.g., \citealt{Burn_1966}; \citealt{Brentjens_2005}; \citealt{Heald_2009}; \citealt{Mao_2010}.) 
We use {\sc rm tools 1d} software
(\citealt{Purcell_2020}) provided by Canadian Initiative for Radio Astronomy Data Analysis (CIRADA).
The idea of RM synthesis is to bring the complex polarised surface brightness $\mathcal{P}$, defined below using Stokes $Q$ and $U$, to the Faraday depth ($\phi$) domain so that one can interpret the changes in $\phi$ along the line-of-sight.
\begin{equation}\label{eq:Stokes_P}
    \mathcal{P}(\lambda^{2})=Q+i\,U.
\end{equation}
A simple approach is to introduce the Faraday dispersion function which is a Fourier conjugate of $\mathcal{P}(\lambda^{2})$.
\begin{equation}
    \mathcal{F}(\phi)={\frac{1}{\pi}}\int_{-\infty}^{\infty} \mathcal{P}(\lambda^{2})e^{-2i\phi\lambda^{2}}d(\lambda^{2}).
\end{equation}
However, there is an incompleteness in $\lambda^{2}$ sampling in the following inevitable reasons:
\begin{itemize}
    \item Mathematically, $\lambda^{2}$ can only be positive.
    \item Stokes $Q$ and $U$, therefore $\mathcal{P}(\lambda^{2})$, are measured within a bandwidth of a telescope and discretized to a finite number of channels.
    \item Some channels are flagged during the data reduction process.
\end{itemize}
The limited sampling of $\lambda^{2}$ space results in sidelobes in ``observed $\mathcal{F}(\phi)$'' (hereafter, $\tilde{\mathcal{F}}(\phi)$) which is distinct from the ideal $\mathcal{F}(\phi)$. The Rotation measure spread function (RMSF), $R(\phi)$ is introduced to describe the discrepancy.
\begin{equation}
    \tilde{\mathcal{F}}(\phi)=\mathcal{F}(\phi)*R(\phi)
\end{equation}

The \texttt{RM-clean} algorithm (\citealt{Heald_2009}) was introduced in order to uncover physically meaningful signal from a dirty Faraday spectrum with noise and sidelobes; a clean Faraday spectrum is obtained by deconvolving the dirty spectrum with an RMSF. The resolution of the Faraday spectra, $\delta\phi$, i.e., FWHM of the RMSF, and the largest detectable scale in $\phi$ space, $\phi_{\rm max-scale}$, are defined as follows:
\begin{equation}
    \delta \phi \approx 2\sqrt{3}/ (\lambda_{\rm max}^{2}-\lambda_{\rm min}^{2}),
\end{equation}
\begin{equation}
    \phi_{\rm max-scale} \approx \pi/\lambda_{\rm min}^{2}.
\end{equation}
Given the bandwidth of our data $0.009\,\rm m^{2}<$ $\lambda^{2}<0.074\,\rm m^{2}$, the estimated FWHM is $\sim53\,\rm rad\,m^{-2}$ and the maximum measurable scale is $\sim349\, \rm rad\,m^{-2}$.
We determined the Faraday depth at the strongest peak in a clean Faraday spectrum and adopted it as the Faraday depth of a source. Therefore, the Faraday depth and RM are interchangeable and the Faraday complexity is not in the scope of this study.

For the catalogue of RM sources used for our analysis, we adopted thresholds as follows.
\begin{enumerate}
    \item The number of channels used as an input for the RM synthesis is larger than 40 so that a certain level of $\lambda^{2}$ coverage is achieved.
    \item The lower limit of $\left\| \phi_{\rm max}\right\|$ is set to $300\,\rm rad\,\rm m^{-2}$ to ensure reasonably small separation between the channels in $\lambda^{2}$ space after flagging (see equation \ref{eq:phi_max}).
    \item The signal-to-noise of the peaks in the Faraday spectra identified using the \texttt{RM-clean} algorithm is larger than 7. 
    \item The observed $\left|\phi\right|$ is smaller than $\left\| \phi_{\rm max} \right\| \approx 750\, \rm rad\,\rm m^{-2}$. This criterion rejects the artificial peaks that appear close to the lower and upper limit of the Faraday spectra.
\end{enumerate}
The resulting 210 sources that match the above criteria are presented in Table \ref{tab:source}.

In order to test the consistency of the observed RMs, we performed a matching between the catalogues. The pairing is based on the sky coordinates of sources.
The top panel of Figure \ref{fig:rm_matching} shows the RMs of matched pairs between ATCA $\times$ NVSS (red) and ATCA $\times$ S-PASS (blue). The size of the symbols corresponds to the angular separation between the matched pairs. The bottom panel presents the histogram of the separation. 
Total 33 and 24 pairs were identified, respectively, with the maximum separation between sources limited to 1 arcmin. We found that most of the matched pairs have approximately identical RMs.

\subsection{S-PASS Galactic diffuse polarisation}

S-PASS provides maps of polarimetric data of the southern sky ($\rm Dec<-1^{\circ}$). The observation of the survey was performed with the S-band ($2.2-3.6\,\rm GHz$) receiver of the Parks radio telescope.
The FWHM of the beam for the final Stokes $I, Q, U,$ and $V$ images is $10.75'$, which is several times enhanced resolution compared with previous continuum surveys of a similar kind (e.g., \citealt{Reich_2001}). This frequency range and the angular resolution enable detailed studies of magnetism at the Galactic disc and the disc-halo interface.
We refer interested readers to \citet{Carretti_2019} for further description of the survey.
In this paper (specifically, Section \ref{sec:diffuse}), we used the Stokes $I$ map to examine the continuum emission at our region of interest and the Stokes $Q$ and $U$ maps to identify depolarised features arising on top of the diffuse polarised emission from the Galactic interstellar medium.

\section{Potential for confusion from sources along the line-of-sight}\label{sec:confusion}

\begin{figure*}
    \centering
    \includegraphics[width=\textwidth]{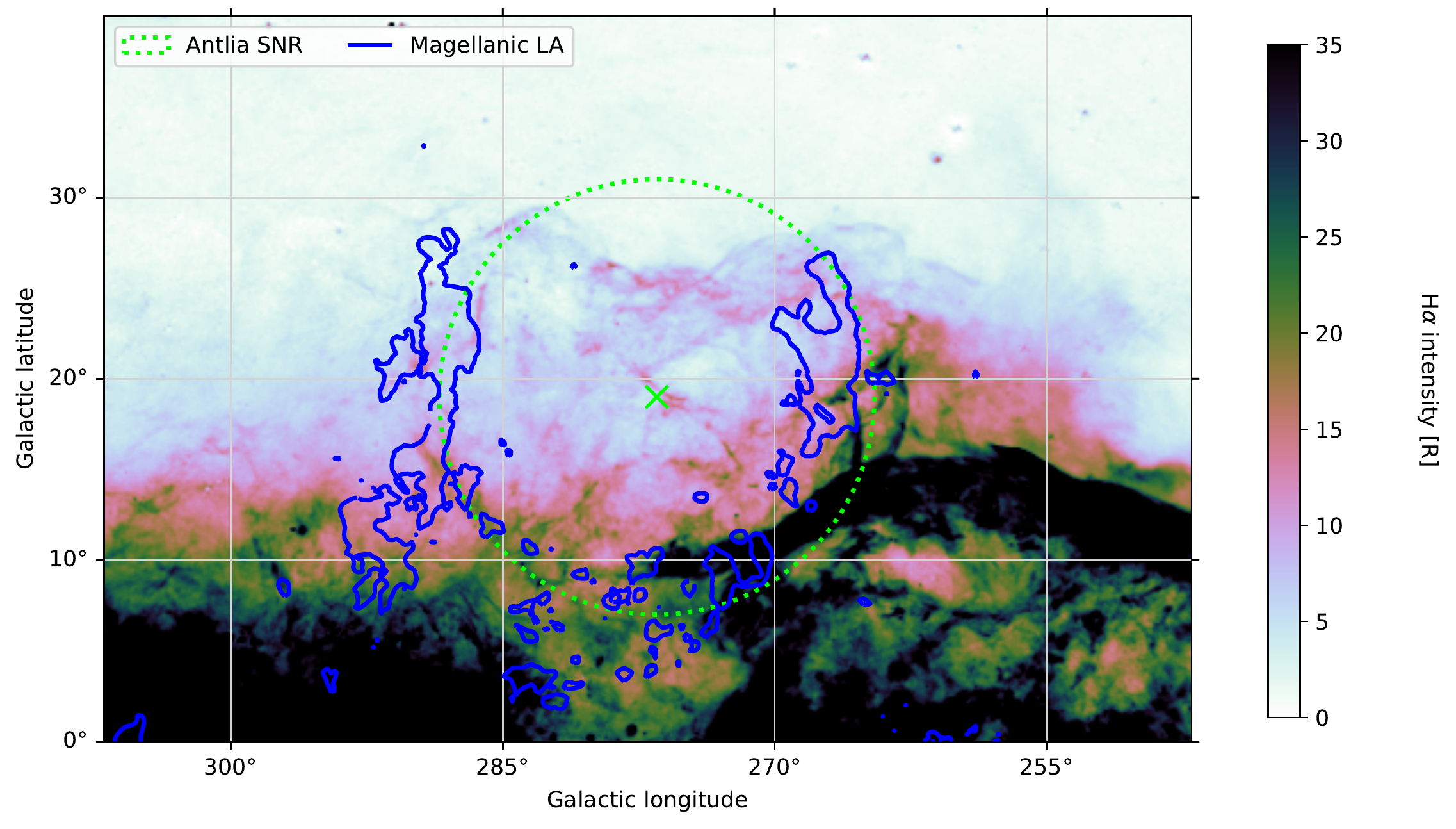}
    \caption{
    The distribution of High-velocity HI clouds ($v_{\rm LSR} = [200-300] \,\rm km/s$; blue contour) overlaid on top of the Galactic $H\alpha$ emission ($v_{\rm LSR} = [-100, 80 ]\,\rm km/s$). 
    The green dotted circle is a schematic drawing of the Antlia SNR with an angular diameter of $24^{\circ}$ (\citealt{McCullough_2002}.)
    }  
    \label{fig:hi+halpha}
\end{figure*}

In this paper, we focus on the Magellanic Leading arm region where the compact-source RM distribution was earlier studied by \citetalias{McClure-Griffiths_2010}.
There are two gaseous objects which are large in solid angle and closely overlap with each other: the Magellanic Leading Arm and the Antlia SNR. Figure \ref{fig:hi+halpha} shows the distribution of high-velocity HI emission (integrated between $200-300\,\rm km/s$; blue contours) from the Galactic All Sky Survey (GASS; \citealt{McClure-Griffiths_2009}) and the Antlia SNR bright in $H\alpha$ composite image at lower velocity range generated by \citet{Finkbeiner_2003} using the Wisconsin H-Alpha Mapper (WHAM; \citealt{Haffner_2003}), the Virginia Tech Spectral line Survey (VTSS), and the Southern $H\alpha$ Sky Survey Atlas (SHASSA; \citealt{Gaustad_2001}). In the following paragraphs, we briefly introduce some known properties of these objects.

The Magellanic Leading Arm is a stream of material tidally stripped out from the Magellanic System during the interaction between the Large Magellanic Cloud, the Small Magellanic Cloud, and the Milky Way (e.g., \citealt{Nidever_2008}; \citealt{Besla_2012}; \citealt{Lucchini_2020}).
A network of large high-velocity complexes, namely, LA $\rm I-IV$, and associated cloudlets are found in HI emission. Their distribution is extended from the Magellanic system to the high latitude sky beyond the Galactic disc. 
In the field studied in this paper, only LA II and LA III are present.

Given the positive galactic latitude of the LA II and the LA III, they are often considered to have passed through the Galactic midplane (\citealt{McClure-Griffiths_2008}).
Recently, a young stellar association was discovered near the tip of the LA II using \textit{Gaia} DR2 (\textit{Price-Whelan 1}; \citealt{Price-Whelan_2019}; \citealt{Nidever_2019}). 
The \textit{PW 1} star cluster is located at $28.7\,\rm kpc$ from the sun. Its estimated age is comparable with the time since the traversing of the Leading Arm through the Galactic disc, which makes the compression of the materials during the interaction a favourable explanation for the formation of the star cluster.

\citetalias{McClure-Griffiths_2010} reported the morphological agreement between the structures in the RM map and the distribution of HVCs in the field.
They suggested that the magnetic field associated with the Leading Arm II is reinforced by relatively strong magnetic fields of the disc that it penetrated before moving into the halo.
Thestrength of the coherent line-of-sight magnetic field  estimated from the RM grid is $B_{\parallel}\gtrsim 6\,\rm \mu G$.
This simple calculation is valid under an assumption that the structures appearing in the RM and the HI emission are physically associated, in other words, if the HVCs are the dominant source of Faraday rotation along the sight-lines.

The Antlia supernova remnant (SNR) is located at $(l,b)=(276.5^{\circ}, +19^{\circ})$ and has a large angular diameter of $24^{\circ}$ (\citealt{McCullough_2002}).
Because such high galactic latitude and size are not common among known Galactic SNRs, whether the object is a supernova-driven remnant or not had been suspected since its first discovery by \citet{McCullough_2002}. Only recently, it is confirmed that the remnant reveals shock-driven emission regions in UV and optical lines supporting the SNR origin (\citealt{Fesen_2021}).
This SNR is bright in $H\alpha$ but weak in radio continuum, suggesting that it is a relatively evolved system.
The distance to the SNR is not well constrained, but the large angular size and features arising from interacting with nearby ISM (e.g., Gum Nebula) locate it around $60-340\,\rm pc$ away within the Galactic disc (\citealt{McCullough_2002}). Note its striking morphological coincidence with the Magellanic Leading Arms presented in Figure \ref{fig:hi+halpha}.

\section{Results} \label{sec:results}

\begin{figure*}
    \centering
    \includegraphics[width=\textwidth]{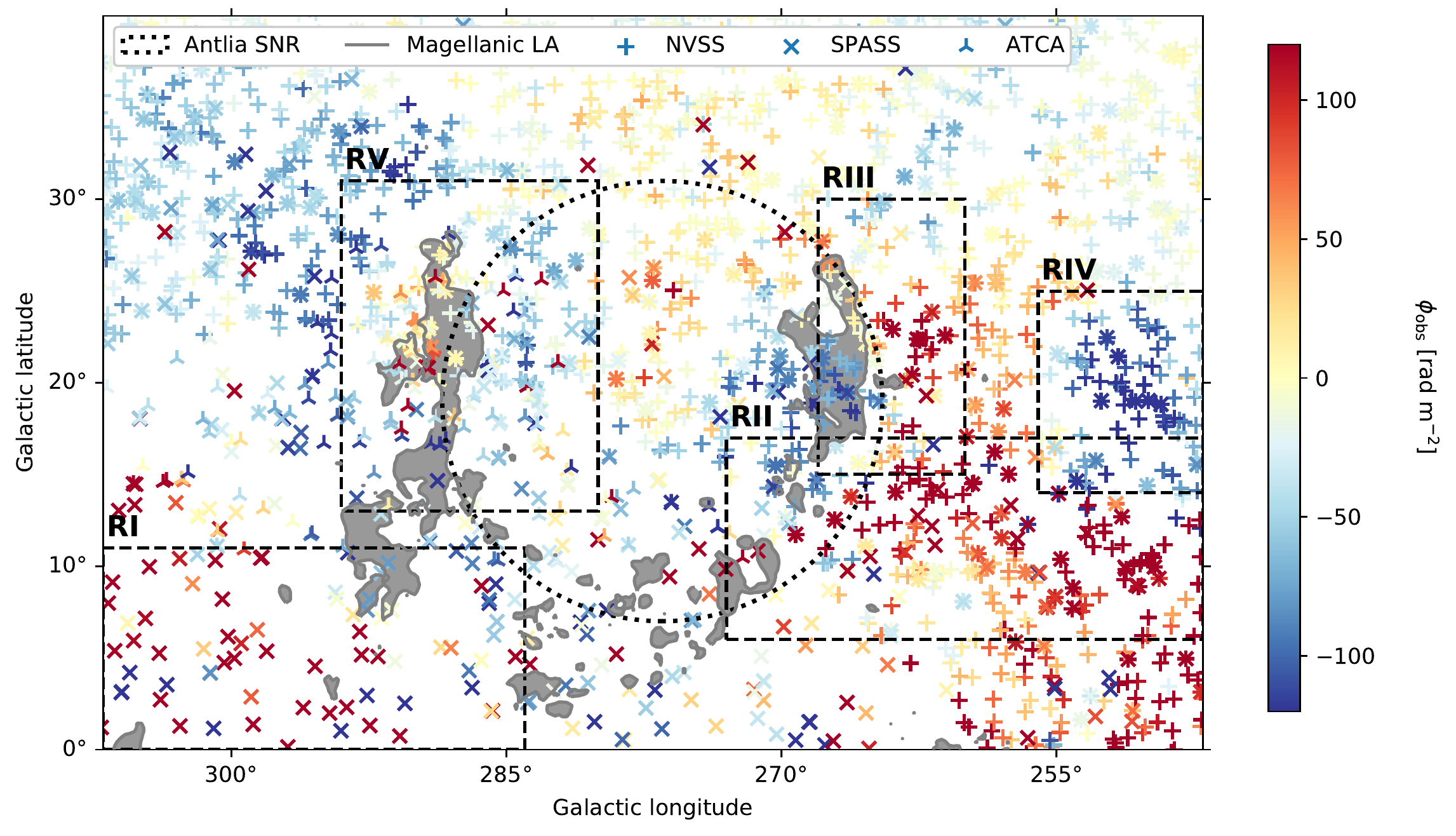}
    \caption{
    Raw RMs from NVSS ($+$ symbol; \citealt{Taylor_2009}), S-PASS/ATCA($\times$ symbol; \citealt{Schnitzeler_2019}), and our observations with ATCA (\protect\rotatebox[origin=c]{180}{$\text{\sffamily Y}$} symbol). The HI contour and the Antila SNR diagram are the same as Figure \ref{fig:hi+halpha}, coloured in black for clarity. The dashed boxes are regions discussed in the text.
    }
    \label{fig:hi+rm}
\end{figure*}

\begin{figure*}
    \centering
    \includegraphics[width=\textwidth]{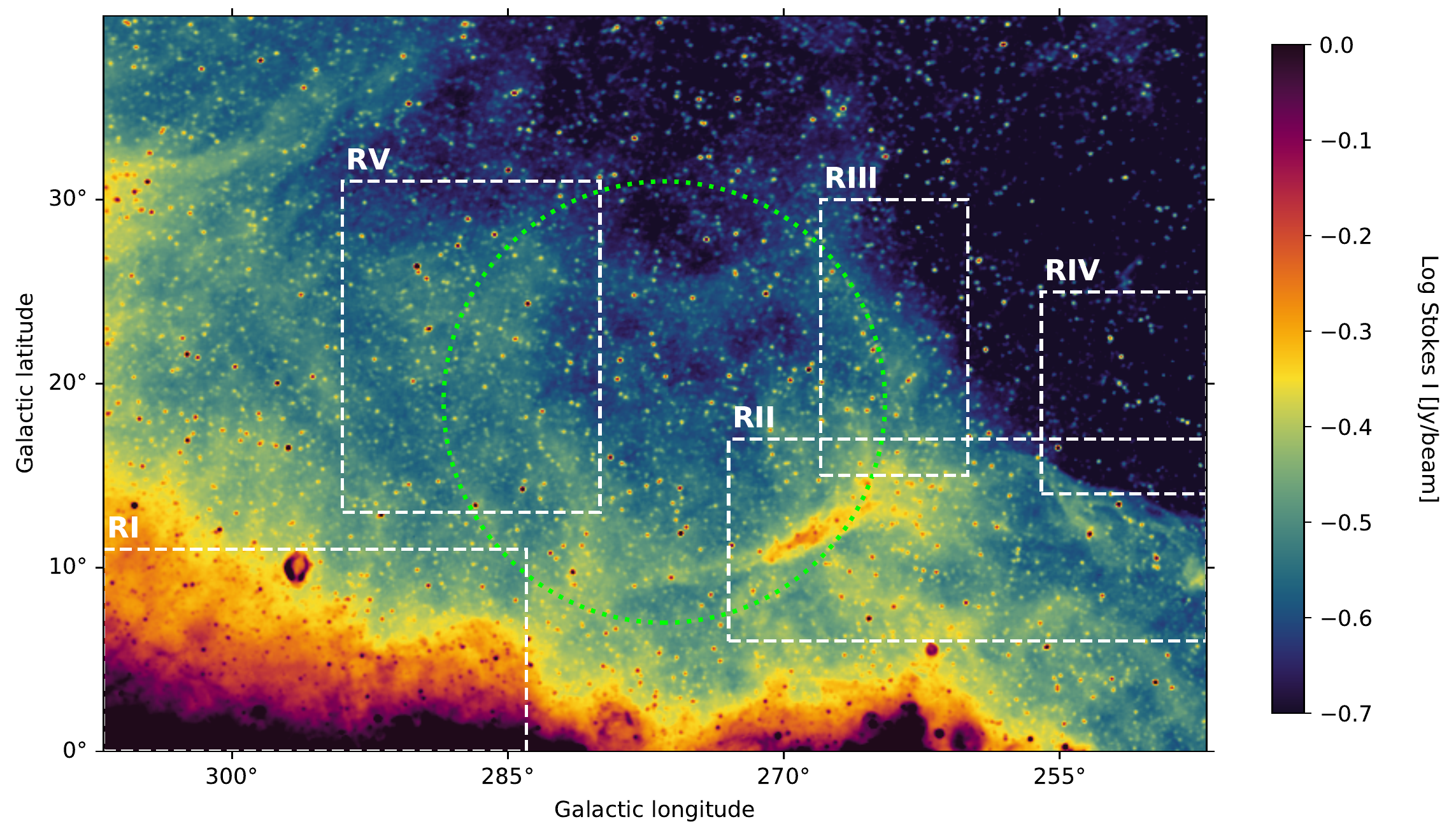}
    \caption{
    The Stokes $I$ image from S-PASS data. The Antlia SNR diagram (green dotted circle) and the white dashed boxes are the same as Figure \ref{fig:hi+rm} above.
    }  
    \label{fig:stokes_i}
\end{figure*}

\begin{figure*}
    \centering
    \includegraphics[width=\textwidth]{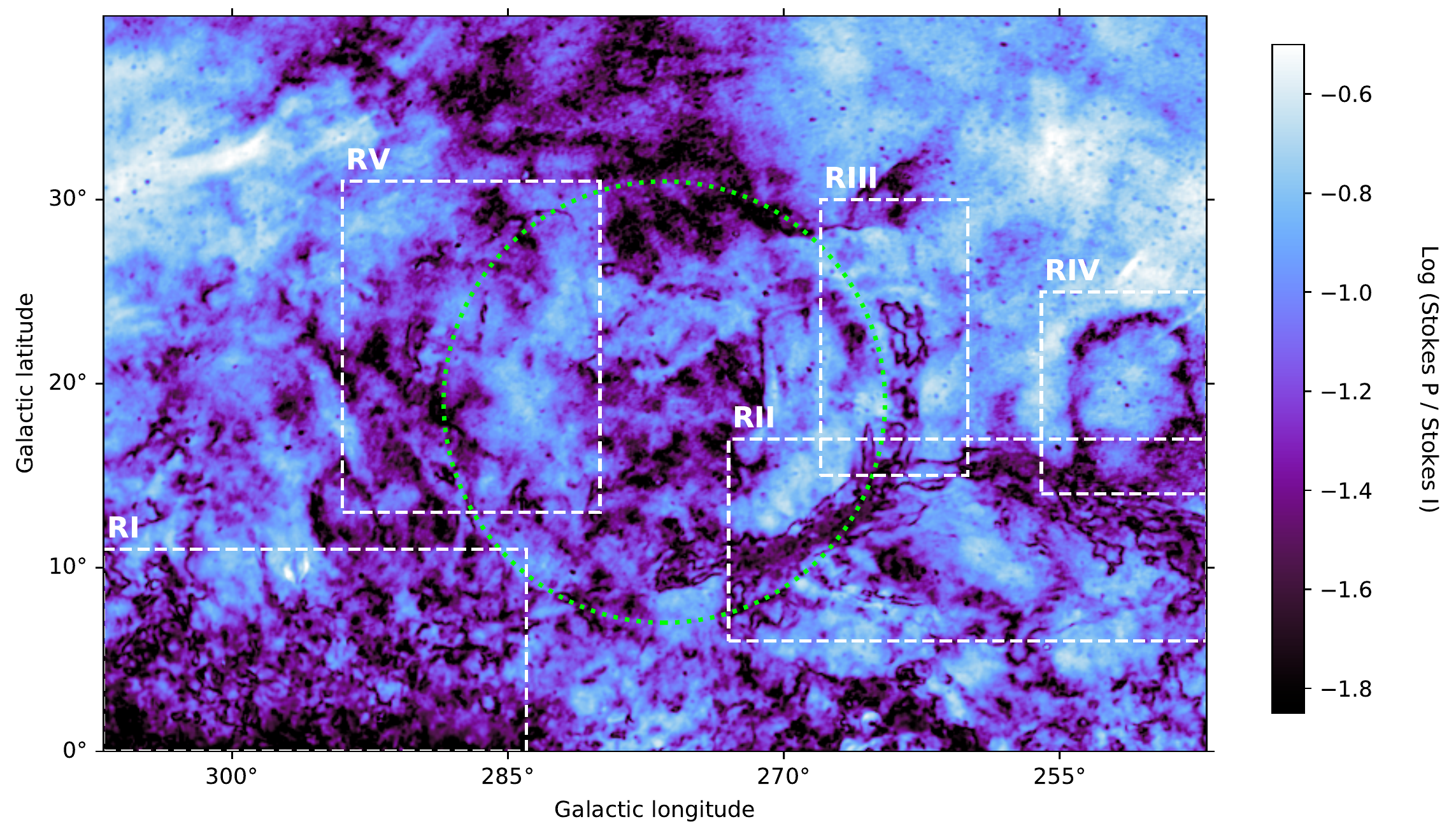}
    \caption{
    The linear polarisation intensity calculated from the Stokes $Q$ and $U$ of S-PASS data.}
    \label{fig:p}
\end{figure*}

\begin{figure*}
    \centering
    \includegraphics[width=\textwidth]{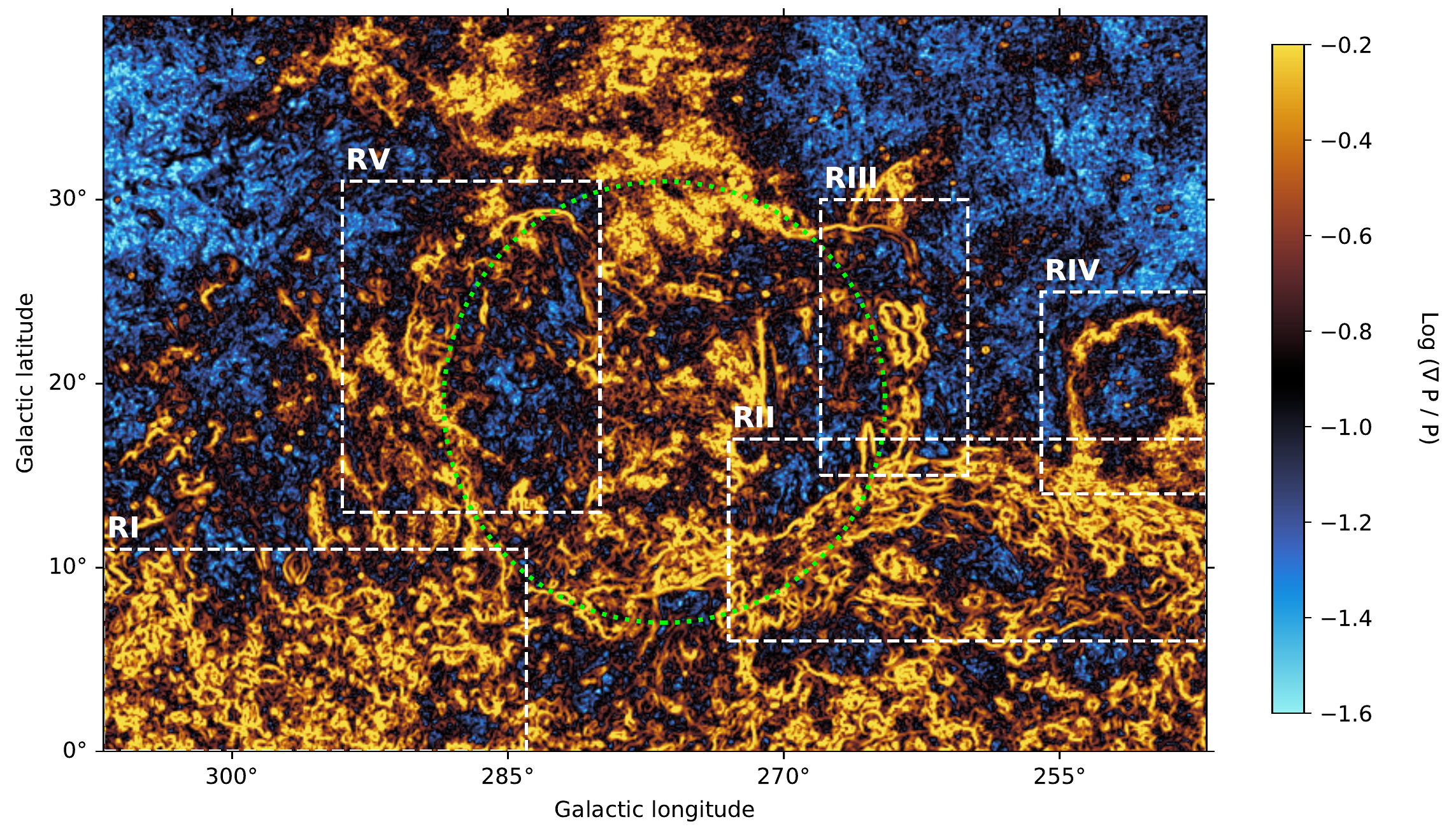}
    \caption{
    The normalized polarisation gradient ($|\nabla \mathcal{P}|/|\mathcal{P}|$) of S-PASS data.
    }
    \label{fig:p_grad}
\end{figure*}

\subsection{Structures in the compact-source RM map} \label{Sec:RM}

To construct the RM grid of the field, we combine 3 RM catalogues as explained in Section \ref{sec:rm_cat}.
We achieved the maximum RM source density of $\approx2\,\rm sources/deg^{2}$. Figure \ref{fig:hi+rm} shows interesting features standing out in the RM grid. We separate the field into several regions labeled in Figure \ref{fig:hi+rm} for convenience in the description.

\begin{enumerate}[start=1,label={Region \Roman*:},wide = 0pt, leftmargin = 4em]
    \item Below galactic latitude $b \lesssim 10^{\circ}$, especially in the south-eastern side of the field, there are large RMs with high fluctuations. 
    The Faraday rotation towards this region is dominated by high density and strongly magnetized ISM cells distributed along the Galactic disc.
    
    \item Toward the south-western corner of the field, there is a group of positive RM sources ($\phi_{\rm obs}>100\,\rm rad\,m^{-2}$) associated with the north-eastern edge of the Gum Nebula which is clearly visible in the $H\alpha$ emission map in Figure \ref{fig:hi+halpha}. \citet{Purcell_2015} constrained the electron density ($n_{\rm e} = 1.4\pm0.4\,\rm cm^{-3}$) and the magnetic field strength ($B=3.9^{+4.9}_{-2.2}\,\rm\mu G$) in this region from the observed RM distribution together with a simple geometric model of a magneto-ionized spherical shell.
    
    \item  There are positive RMs extending along $l\sim 265^{\circ}$ up to $b\sim 25^{\circ}$. 
    This pillar of positive RMs seems to extend from the RM distribution of the Gum Nebula in Region II, but also aligns well with the western edge of the Antlia SNR and LA III. Also, it was earlier reported by \citet{Reynoso_1997} that there is a vertical HI structure in this region that is possibly related with blown-out ISM from the Galactic disc to the halo (i.e., a galactic chimney).
    Due to the complexity in the region, it is difficult to identify which object is a dominant Faraday rotator along the line-of-sight.
    
    \item We also notice a group of negative RMs ($\phi_{\rm obs}<-100\,\rm rad\,m^{-2}$) near the western boundary of the field. Interestingly, there is no corresponding radio continuum nor $H\alpha$ emission detected nor previously reported magneto-ionized object in this region.
    
    \item At the eastern edge of the Antlia SNR where it overlaps with the LA II, there is a group of RMs close to zero surrounded by negative RMs. This is the region studied by \citetalias{McClure-Griffiths_2010}. \citetalias{McClure-Griffiths_2010} proposed the RM excess in the region as indication of magnetized LA II, but the Antlia SNR in the foreground complicates the determination of the dominant Faraday rotator in the region.
\end{enumerate}

Our inspection of the noticeable structures in the RM grid indicates that it is not uncommon to find multiple gaseous objects along the lines-of-sights. Especially, the situation is complicated in Region III and Region V due to the coincidence of the Antlia SNR and the Magellanic Leading Arms.
Additional pieces of information is required in order to determine where along the lines-of-sight the Faraday rotation occurs in this region.

\subsection{Radio continuum and diffuse polarised emission}\label{sec:diffuse}

\begin{figure}
    \centering
    \includegraphics[width=\columnwidth]{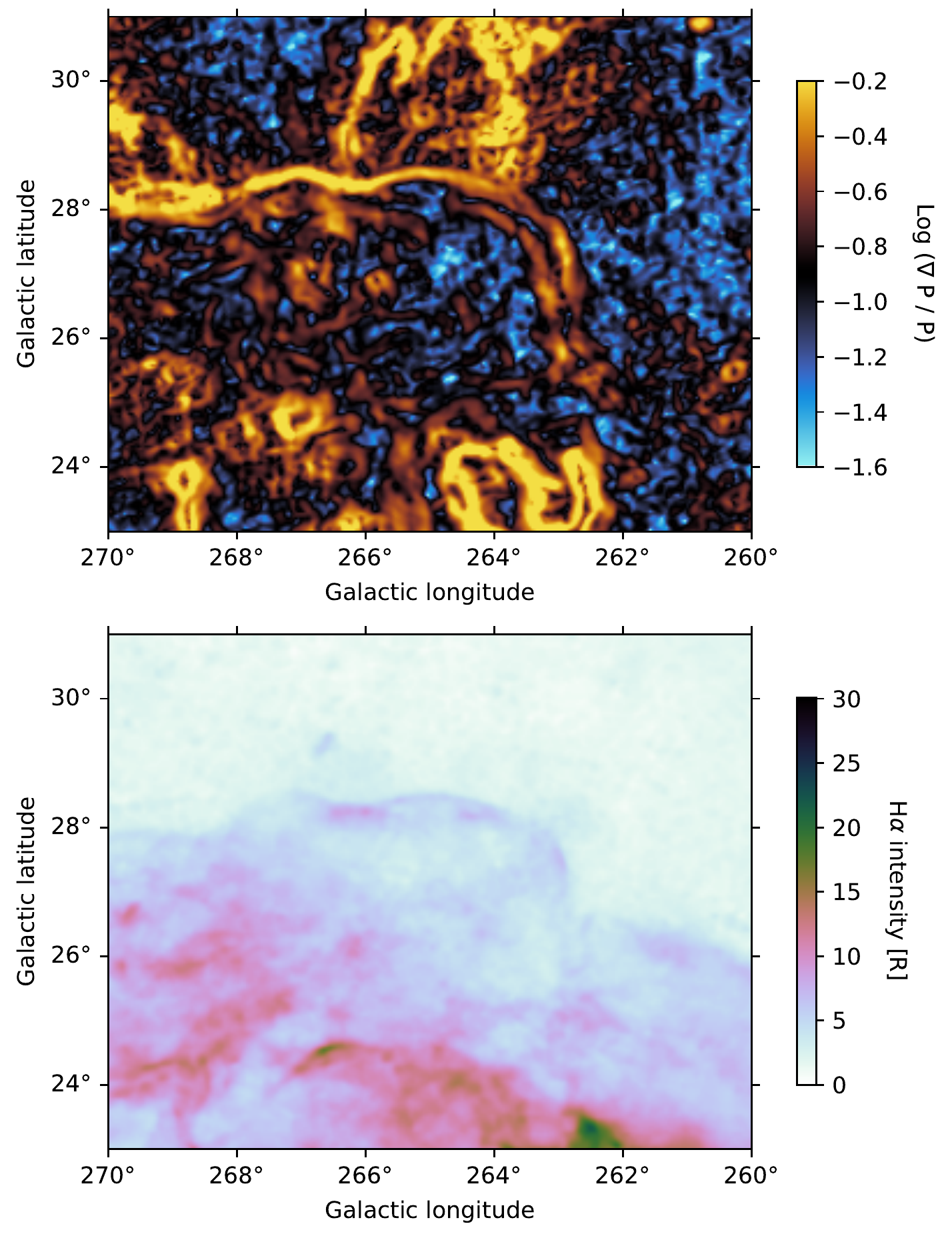}
    \caption{
    The zoom-in image of a $|\nabla \mathcal{P}|/|\mathcal{P}|$ double-jump profile filament in Region III (the north-western edge of the Antlia SNR) and its $H\alpha$ counterpart.
    }
    \label{fig:zoom-in}
\end{figure}

Figure \ref{fig:stokes_i} shows the S-PASS Stokes $I$ image of the field\footnote{We used colour maps provided by {\sc CMASHER} python package (\citealt{cmasher}) in Figure \ref{fig:stokes_i}, \ref{fig:p}, and \ref{fig:p_grad}.}. 
Our field of interest covers a wide range of synchrotron emitting regions including the bright Galactic disc (e.g., Region I) as well as the lower-brightness diffuse structures. 
There is weak radio synchrotron emission in the Antlia SNR region. However, its continuum emission is not as prominent as in $H\alpha$ except for the south western edge of the SNR where it interacts with the Gum Nebula (Region II).
In comparison to the Stokes $I$ image, the linear polarisation intensity map (Figure \ref{fig:p}) shows rich filamentary structures on top of the smooth polarised emission.
These depolarised filaments appear if there is a sharp change in electron and/or magnetic properties in the foreground.

The properties of linearly polarised radiation are often characterised with the complex Stokes vector $\mathcal{P}$, as defined in equation \ref{eq:Stokes_P} (\citealt{Gaensler_2011}). 
Assuming the spatial gradient in $\mathcal{P}$, i.e.,
\begin{equation}
    |\nabla \mathcal{P}| = \sqrt{\left( \frac{\partial Q}{\partial x}\right)^{2} + \left( \frac{\partial U}{\partial x}\right)^{2} + \left( \frac{\partial Q}{\partial y}\right)^{2} + \left( \frac{\partial U}{\partial y}\right)^{2}},
\end{equation}
arises from fluctuations in the Faraday rotation in foreground Faraday screens, we adopt the normalized parameter, $|\nabla \mathcal{P}|/|\mathcal{P}|$ (Figure \ref{fig:p_grad}). Taking the spatial gradient makes it easier to trace edges of the filaments standing out in Figure \ref{fig:p}. Bright features in the $|\nabla \mathcal{P}|/|\mathcal{P}|$ map therefore highlight depolarisation arising from complex layers of Faraday screens in this region of the sky.


With the normalized polarisation gradient map in Figure \ref{fig:p_grad}, we revisit the regions discussed earlier in Section \ref{Sec:RM} where we identify noticeable distinction in the distributions of RMs. 
\begin{enumerate}[start=1,label={Region \Roman*},wide = 0pt, leftmargin = 4em]
    \item (bottom left): This area is covered with chaotic small angular scale depolarised filaments. These fuzzy structures are typical at low galactic latitude (e.g., \citealt{Uyaniker_2003}; \citealt{Iacobelli_2014}) where significant depolarisation is expected from the turbulent ISM in the Galactic disc.
    
    \item (bottom right): The bright web of depolarised filaments spread along the north-eastern edge of the Gum Nebula indicates that the object effectively operates as a Faraday rotating screen.
    This is not surprising given its (i) high emissivity in $H\alpha$ (see Figure \ref{fig:hi+halpha}) suggesting the presence of a substantial amount of free electrons and (ii) the morphological coherence in the RM grid indicating that the nebula is magnetized.
    
    \item (right edge of the dotted circle): There are depolarised filaments and loops extended along the western edge of the Antlia SNR. The remarkable spatial coherence of $H\alpha$ emission and the $|\nabla \mathcal{P}|/|\mathcal{P}|$ distribution in this region was earlier reported by \citet{Iacobelli_2014}.
    Furthermore, we discovered a narrow filament with ``double-jump'' profile at $(l,b) \approx (263, 27)$ which indicates a delta function-like distribution of $n_{\rm e}$ and/or magnetic field (e.g., a strong shock; \citealt{Burkhart_2012}). Indeed, \citet{Fesen_2021} studied the spectral line ratios of the very filament and supported the shock origin of the filament. We present a zoom-in image of this filament in Figure \ref{fig:zoom-in}.
    
    \item (middle right): The striking feature standing out in this region is a narrow loop. 
    Its single-jump profile suggests a step-function-like change in magnetic field properties in and outside the loop (\citealt{Burkhart_2012}). 
    The negative RMs found in Figure \ref{fig:hi+rm} are well enclosed within the loop, indicating the enhanced magnetic field strength in this region.
    
    \item (left edge of the dotted circle): 
    Similarly to Region III, there is a complex network of filaments in this region including a narrow filament at $(l,b) \approx (284,29)$ that clearly overlap with a narrow $H\alpha$-emitting filament. 
\end{enumerate}


\section{Discussion}\label{sec:discussion}

\subsection{Is the Antlia SNR a Faraday rotator?}
\begin{figure*}
    \centering
    \includegraphics[width=\textwidth]{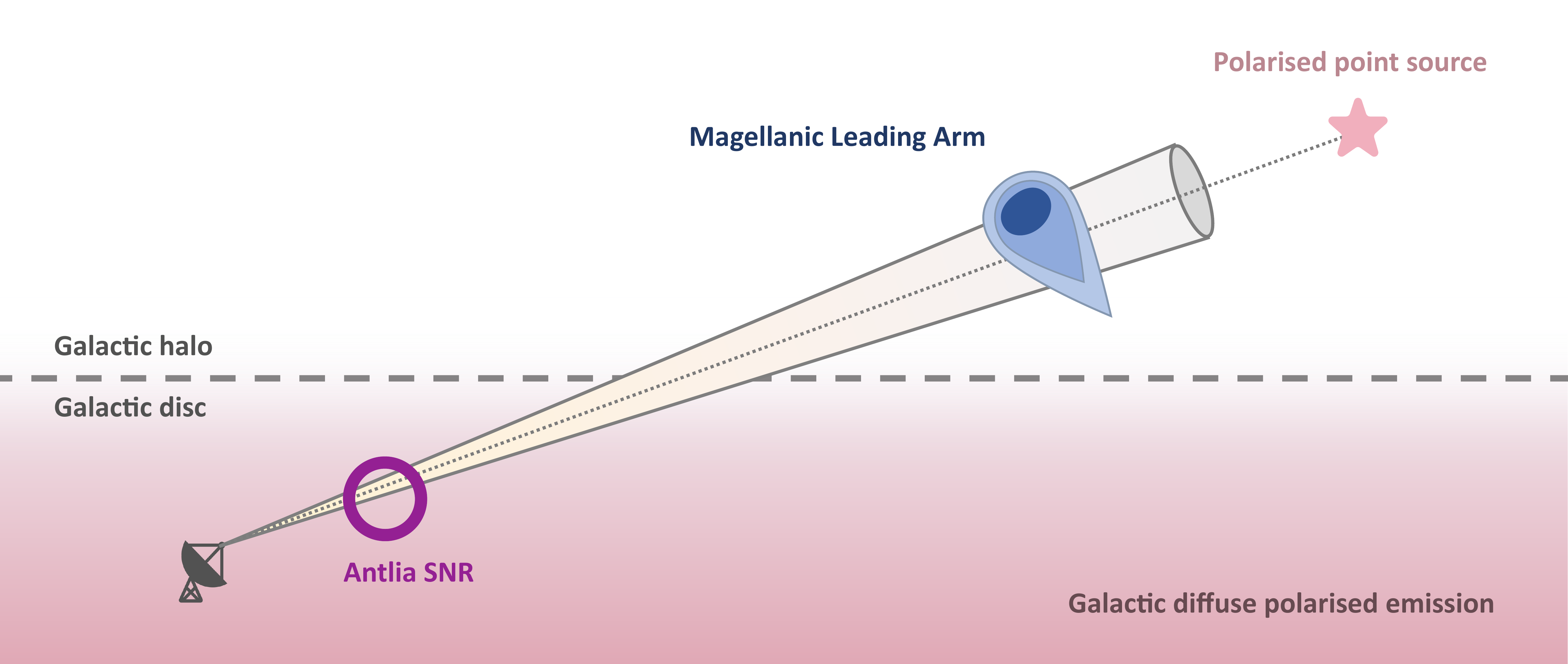}
    \caption{Illustration of the Antlia SNR/Magellanic Leading Arm field. The polarised radiation from the extra galactic point source propagate through both the Magellanic Leading Arm and the Antlia SNR, while the Galactic diffuse polarised emission knows only about the Antlia SNR. This image is for illustration purpose only. The size of and the distance to the objects does not correspond to the real size and the distance.
    }  
    \label{fig:illust}
\end{figure*}


The observed RM of a distant source ($\phi_{\rm obs}$) is a superposition of the Faraday rotation occurring at every magneto-ionized medium between the source and the observer. This includes the Faraday rotation at the polarised emitting source ($\phi_{\rm intrinsic}$) and the Milky Way foreground ($\phi_{\rm MW}$): 
\begin{equation}\label{eq:phi_obs}
    \phi_{\rm obs} = \phi_{\rm intrinsic}+\sum_{i=1}^{N}\phi_{\rm obj,\, i}+\phi_{\rm MW},
\end{equation}
where $\phi_{\rm obj, i}$ represents the Faraday rotation taking place at different distance along the line-of-sight and $N$ is the number of such Faraday screens which is very likely unknown for any sight-lines.
RM catalogues from all-sky surveys revealed large Galactic-scale structures in the RM grids which indicates that $\phi_{\rm MW}$ is likely to dominate the observed RMs in most of the sky (\citealt{Taylor_2009}; \citealt{Schnitzeler_2019}).
In studies of objects with smaller angular scales like HVCs, the contribution of $\phi_{\rm MW}$ is often estimated using off-object RMs of the region and subtracted from $\phi_{\rm obs}$.
The variation of the intrinsic polarisation of sources ($\phi_{\rm intrinsic}$) is random and therefore negligible on the basis of the large number statistics.

In the region of the sky studied in this paper, there are several known localized objects that could possibly induce Faraday rotation ($\phi_{\rm obj, i}$), if magnetized: the Gum Nebula, the Antlia SNR, and the Magellanic Leading Arm. 
The Gum Nebula leaves an imprint on the RM grid that closely follows the morphology of its $H\alpha$ emission (see Region II in Figure \ref{fig:hi+rm}), making it clear that it is the dominating Faraday rotator in the region.
However, the overlap of the Antlia SNR and the Magellanic Leading Arm on the sky makes it indeterminate whether the features in the RM grid arise due to either of or both the objects.
They are not physically associated given their distinct observed velocities and distances.
Therefore, if the Antlia SNR is magnetized and significantly affects the RM towards the sightlines, the RM signiture identified by \citetalias{McClure-Griffiths_2010} can no longer be clearly associated with the Leading Arm.

To test the possibility of whether the intriguing features appearing in the compact-source RM map are associated with the Magellanic Leading Arm in the Galactic halo or the Antlia SNR in the foreground, we bring extra information from the diffuse polarised radiation emitted from the large-scale Galactic interstellar medium.
The diffuse polarised emission traces large-scale smooth polarised emission from Galactic ISM.
On top of that, Faraday screens that alter the polarisation properties of radiation coming from behind produce depolarised structures.
When polarised radiation emitted from different patches of ISM are combined into a beam, their properties are inevitably averaged out due to the turbulent nature of ISM (i.e. large fluctuations in properties) and finite beam sizes in radio observations.
This depolarisation effect result in iconic filamentary structures in a polarisation emission, e.g., depolarised canals (e.g., \citealt{Haverkorn_2004}; \citealt{Fletcher_2006}).

Figure \ref{fig:illust} illustrates the relative location of the objects and roughly where the polarised radiation of point-sources and the Galactic diffuse emission come from with respect to the objects.
Unlike extragalactic radiation that propagate through both the Magellanic Leading Arm and the Antlia SNR, the Galactic diffuse polarisation does not experience Faraday rotation (if there is any) at the Magellanic Leading Arm since the object is beyond the Galactic disc where most of the emission is coming from.
Therefore, the morphological correspondence between the low-velocity $H\alpha$ filaments (Figure \ref{fig:hi+halpha}) and the depolarised canals (Figure \ref{fig:p} and Figure \ref{fig:p_grad}) is the smoking gun evidence that the Antlia SNR is a Faraday rotator and severely affects the observed RM towards the region where the excess of RM was reported by \citetalias{McClure-Griffiths_2010}.
Our findings lead to a conclusion that it is not feasible to draw any certain conclusions about the magnetic fields of the Magellanic Leading Arm using the RM grid technique. In other words, \emph{it is hard to interpret the RM excess in this region as evidence of the ``magnetized'' Leading Arm}


We attempt to perform a similar test on the Smith cloud, which is another candidate of magnetized HVCs, but it was impossible since the cloud is located at the celestial equator which is right at the border of the S-PASS sky coverage. To our knowledge, there is no comparable polarimetric survey at the same frequency that covers the northern sky.
Yet, the low-velocity H$\alpha$ emission in the region does not show any structures above $5\,\rm Rayleigh$ unlike the case of the Magellanic Leading Arm. We conclude that the Smith Cloud region has a less chance of being affected by the Galactic foreground.


\subsection{Magnetic field strength of the Antlia SNR}\label{sec:b_los}

\begin{figure}
    \centering
    \includegraphics[width=\columnwidth]{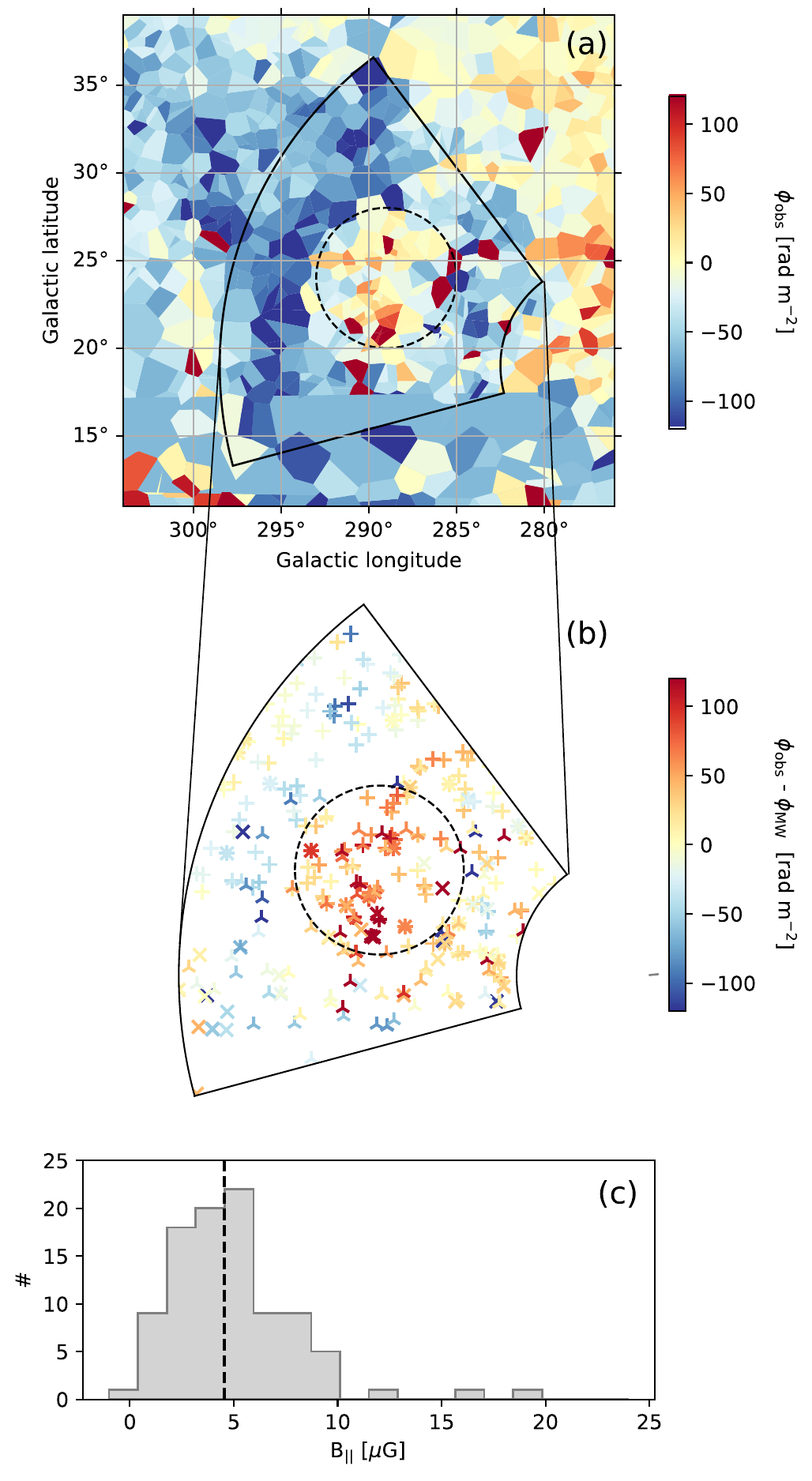}
    \caption{
    Panel (a): A Voronoi diagram coloured with $\phi_{\rm obs}$. The RM excess is enclosed within a dashed circle.
    Panel (b): the Milky Way corrected RM ($\phi_{\rm obs}-\phi_{\rm MW}$) in the RM excess region and the surrounding.
    Panel (c): the distribution of $B_{\parallel}$ estimated from sources in the RM excess region
    }
    \label{fig:b_los}
\end{figure}

In this Section, we calculate the line-of-sight magnetic field strength at the eastern edge of the Antlia SNR (i.e., Region V of Figure \ref{fig:hi+rm}, \ref{fig:stokes_i}, \ref{fig:p}, and \ref{fig:p_grad}) assuming the RM excess is entirely due to the magneto-ionized shell of the Antlia SNR. 
This is to check if the estimated field strength is in a reasonable range expected from typical SNRs, in other words, whether the observed RM excess can be explained solely by Faraday rotation at the Antlia SNR.

From the definition of the RM (equation \ref{eq:RM}), the line-of-sight magnetic field strength can be expressed as follows:
\begin{equation}\label{eq:B}
    B_{\parallel} = \frac{\phi_{\rm SNR}}{0.812  \left<n_{\rm e}\right>   f L},
\end{equation}
where $\phi_{\rm SNR}$ is RM of the SNR, $f$ is a volume filling factor of the ionized gas set to 0.5 and $L$ is a path length .
Note that, here we work with the product of an average electron density and the path length through the SNR, $\left<n_{\rm e}\right> L$, since the distribution of the electron density along the sight-line, $n_{\rm e}(r)$, is not known.
Therefore, $B_{\parallel}$ is by assumption the electron-density weighted average magnetic field strength along the line-of-sight.

Under the assumption that the SNR is a dominant source of Faraday rotation other than the smoothly varying large-scale Milky Way ISM, equation \ref{eq:phi_obs} can be expressed as
\begin{equation}
\phi_{\rm SNR}=\phi_{\rm obs}-\phi_{\rm MW}
\end{equation}
in the region where the excess of RM is.
The Voronoi diagram in panel (a) of Figure \ref{fig:b_los} shows the overall distribution of $\phi_{\rm obs}$ near the eastern edge of the Antlia SNR.
The RM excess region is enclosed with a dashed circle.

On the other hand, in the surrounding domain where the RM distribution does not show any correspondence with the SNR, we can consider that the Milky Way is the only Faraday screen (i.e., $\phi_{\rm obs}\approx\phi_{\rm MW}$).
We estimate $\phi_{\rm MW}$ of the RM excess region based on $\phi_{\rm obs}$ measurements in the surrounding.
This approximation holds only at a immediate vicinity where the variance in the large-scale Milky Way field is small and there is no other apparent Faraday screens (e.g., the Gum nebula). Therefore, we restrict our sample to sources within a region bounded by the solid line in Panel (a).
The median $\phi_{\rm obs}$ in this region is adopted as $\phi_{\rm MW}=-56.55\,\rm rad\,m^{-2}$.
Panel (b) shows the distribution of $\phi_{\rm obs}-\phi_{\rm MW}$.
The corrected RMs are mostly positive within the RM excess region as previously pointed out by \citetalias{McClure-Griffiths_2010} and nearly zero in the surrounding region.

The path length through the SNR ($L$) is estimated from a simple geometric model of a 3D spherical shell.
We define $L$ as a function of the angular separation ($\theta$) from the center of the SNR at $(l, b) = (276.5^{\circ}, +19^{\circ})$:
\begin{equation}
    L(\theta) = 
    \begin{cases}
         2D\sqrt{\sin^{2}\theta_{\rm shell}-\sin^{2}{\theta}} &\text{if $\theta_{\rm shell} - d\theta<\theta<\theta_{\rm shell}$}\\
         0 &\text{if $\theta>\theta_{\rm shell}$}\\
         2 D d\theta &\text{if $\theta<\theta_{\rm shell} - d\theta$}
    \end{cases},
\end{equation}
where $D=100\,\rm pc$ is the distance to the Antlia SNR, $\theta_{\rm shell}=18^{\circ}$ and $d\theta=8^{\circ}$ are the outer radius and the thickness of the shell, respectively, in angular scale. 

The average electron density along the sight-lines ($\left<n_{\rm e}\right>$) is estimated from the emission measure ($EM$) of the SNR from its $H\alpha$ intensity:
\begin{equation}
    EM = 2.75\left(\frac{T_{\rm e}}{10^{4}\rm K}\right)^{0.9} I_{H\alpha}, 
\end{equation}
where $T_{\rm e}=10^{4}\,\rm K$ is the electron temperature and $I_{H\alpha}$ is the $H\alpha$ emission in Rayleighs. From the definition of $EM$,
\begin{equation}
    \left<n_{\rm e}\right> = \sqrt{\frac{EM}{f L}}.
\end{equation}

All parameters combined, equation \ref{eq:B} can be expressed as
\begin{equation}
    B_{\parallel} = \frac{\phi_{\rm obs}-\phi_{\rm MW}}{0.673\sqrt{I_{H\alpha}} L }.
\end{equation}
The histogram in Panel (c) shows the distribution of $B_{\parallel}$ calculated using the individual polarised sources in the RM excess region.
The median of the distribution is at $B_{\parallel}\approx 5\,\rm\mu G$, which is similar to a typical magnetic field strength in the Galactic ISM.

\section{Summary}\label{sec:summary}

We hereby summarise three mutually related major points discussed throughout this paper.
First, we argue that the Faraday rotation towards the Magellanic Leading Arm is highly affected by the foreground SNR. 
\citetalias{McClure-Griffiths_2010} identified the Magellanic Leading Arm as a magnetized HVC. However, we find that the Antlia SNR can equally well explain the observed RM excess.

Second, our work provides information about the structures and the magnitude of magnetic fields associated with the Antlia SNR.
The remnant is a Faraday rotator that severely depolarises the diffuse Galactic polarised emission in the background.
We also found double-jump profile filaments in the normalized polarisation gradient map which indicate a sharp enhancement along the shock regions that are bright in $H\alpha$ emission and studied by \citet{Fesen_2021}.
From the compact source RM grid, we estimated the line-of-sight magnetic field strength at the Eastern edge of the Antlia SNR to be $B_{\parallel}\approx5\,\rm\mu G$.

Finally, the lesson we learned from our study in the Antlia SNR/ Magellanic Leading Arm field raise caution for future studies using the RM grid technique.
Up-coming radio telescopes and polarisation all-sky surveys are expected to significantly increase the RM source density. This will allow us to study magnetism in much detail and even extend the RM grid technique to extra galactic objects (e.g. \citealt{Anderson_2021}).
However, such studies should always be aware that there are local Faraday rotators that can significantly affect the RM grid, like the Antlia SNR in our case. We suggest checking Galactic diffuse polarisation maps to identify local Faraday screens.

\section*{Acknowledgements}

We thank the anonymous referee for the positive and constructive review of the work. 
The Australia Telescope Compact Array is part of the Australia Telescope National Facility which is funded by the Australian Government for operation as a National Facility managed by CSIRO. We acknowledge the Gomeroi people as the traditional owners of the Observatory site.
We also acknowledge the Ngunnawal and Ngambri people as the traditional owners and ongoing custodians of the land on which the Research School of Astronomy \& Astrophysics is sited at Mt Stromlo.
N.M.G. acknowledges the support of the ARC through Future Fellowship FT150100024.
A.S.H. is partly supported by a Discovery Grant from the Natural Sciences and Engineering Research Council of Canada.

Our analysis was performed using the Python programming language (Python Software Foundation, https://www.python.org). The following packages were used throughout the analysis: numpy (\citealt{Harris_2020}),  SciPy (\citealt{Virtanen_2020}), and matplotlib (\citealt{Hunter_2007}). This research additionally made use of the publicly available tools: MIRIAD (\citealt{Sault_1995}), RM tools 1D (\citealt{Purcell_2020}), and {\sc CMASHER} python package (\citealt{cmasher}).

\section*{Data availability}
The data underlying this article were accessed from the CSIRO Australia Telescope National Facility online archive at https://atoa.atnf.csiro.au, under the project codes C2741. The derived data generated in this research are available in the article and in its online supplementary material.




\bibliographystyle{mnras}
\bibliography{references} 

\appendix
\section{Table of RM sources}

\onecolumn
\begin{longtable}{ccccccccccc}
\caption{This table includes information about polarised sources observed using ATCA. See Section \ref{sec:atca} for details of observations and how they are identified.}\label{tab:source}\\
\hline
R.A. & Dec. & l & b & $S_{\rm I}$ & $S_{\rm I, err}$ & $p$ & $p_{\rm err}$ & $\phi$ & $\phi_{\rm err}$ \\
(J2000) & (J2000) & ($^{\circ}$) & ($^{\circ}$) & (Jy/beam) & (Jy/beam) & (Jy/beam) & (Jy/beam) & (rad m$^{-2}$) & (rad m$^{-2}$) \\
\hline
\hline
\endfirsthead
\multicolumn{8}{@{}l}{(continued)}\\
\hline
R.A. & Dec. & l & b & $S_{\rm I}$ & $S_{\rm I, err}$ & $p$ & $p_{\rm err}$ & $\phi$ & $\phi_{\rm err}$ \\
(J2000) & (J2000) & ($^{\circ}$) & ($^{\circ}$) & (Jy/beam) & (Jy/beam) & (Jy/beam) & (Jy/beam) & (rad m$^{-2}$) & (rad m$^{-2}$) \\
\hline
\hline
\endhead\\
\endfoot
\hline
\endlastfoot
11h21m22.9s & -43d55m37.9s & 286.2590 & 16.0097 & 0.08 & 0.02 & 0.03 & 0.002 & -36.97 & 2.21 \\
11h16m58.9s & -43d44m53.9s & 285.4225 & 15.8887 & 4.87 & 0.04 & 0.70 & 0.003 & -61.41 & 0.15 \\
11h14m41.3s & -41d09m25.5s & 283.9909 & 18.1291 & 6.97 & 0.08 & 0.99 & 0.006 & -59.14 & 0.26 \\
11h14m39.5s & -41d09m14.3s & 283.9840 & 18.1298 & 7.32 & 0.07 & 1.43 & 0.007 & -80.34 & 0.20 \\
11h10m39.2s & -41d18m48.2s & 283.3210 & 17.6888 & 2.29 & 0.01 & 0.04 & 0.001 & -27.11 & 0.90 \\
11h20m19.0s & -38d22m39.7s & 283.9437 & 21.1111 & 5.47 & 0.12 & 0.43 & 0.013 & -106.20 & 1.27 \\
11h24m05.5s & -36d48m32.7s & 284.0781 & 22.8442 & 0.94 & 0.04 & 0.06 & 0.002 & -51.68 & 1.30 \\
11h16m16.8s & -37d00m07.9s & 282.5887 & 22.0820 & 0.64 & 0.01 & 0.05 & 0.002 & -4.95 & 1.18 \\
11h12m36.9s & -37d45m46.0s & 282.1902 & 21.0965 & 0.98 & 0.12 & 0.05 & 0.011 & 209.59 & 10.09 \\
11h14m41.3s & -41d09m25.6s & 283.9906 & 18.1290 & 6.50 & 0.08 & 0.65 & 0.004 & -78.47 & 0.23 \\
11h14m39.5s & -41d09m15.6s & 283.9841 & 18.1294 & 5.46 & 0.04 & 0.78 & 0.003 & -81.10 & 0.17 \\
11h10m39.3s & -41d18m49.2s & 283.3214 & 17.6886 & 1.60 & 0.01 & 0.09 & 0.001 & -30.93 & 0.52 \\
11h10m33.8s & -41d18m16.2s & 283.3010 & 17.6902 & 0.87 & 0.01 & 0.12 & 0.001 & -35.44 & 0.33 \\
11h29m33.4s & -42d49m15.6s & 287.3244 & 17.5556 & 2.23 & 0.03 & 0.22 & 0.003 & -56.40 & 0.51 \\
11h38m12.9s & -42d45m56.8s & 288.8927 & 18.1010 & 2.20 & 0.07 & 0.25 & 0.005 & -41.58 & 0.82 \\
11h38m12.9s & -42d45m56.4s & 288.8927 & 18.1011 & 1.93 & 0.03 & 0.33 & 0.001 & -33.52 & 0.18 \\
11h38m12.9s & -42d45m56.2s & 288.8925 & 18.1012 & 2.09 & 0.10 & 0.14 & 0.005 & -42.17 & 1.58 \\
11h40m11.2s & -40d49m09.7s & 288.6633 & 20.0699 & 2.07 & 0.02 & 0.04 & 0.001 & -5.35 & 1.55 \\
11h28m14.8s & -39d08m32.3s & 285.7825 & 20.9431 & 3.57 & 0.02 & 0.18 & 0.002 & -46.09 & 0.35 \\
11h24m18.0s & -39d49m06.2s & 285.2681 & 20.0458 & 1.99 & 0.03 & 0.22 & 0.003 & -47.12 & 0.65 \\
11h20m13.6s & -41d54m12.0s & 285.2944 & 17.8233 & 0.51 & 0.04 & 0.18 & 0.004 & -3.17 & 0.79 \\
11h37m06.1s & -44d11m01.8s & 289.1287 & 16.6864 & 0.67 & 0.03 & 0.14 & 0.005 & -137.23 & 1.27 \\
11h34m26.6s & -44d07m48.7s & 288.6374 & 16.5941 & 2.24 & 0.04 & 0.48 & 0.002 & -135.57 & 0.19 \\
11h34m24.6s & -44d06m42.7s & 288.6258 & 16.6097 & 1.19 & 0.04 & 0.24 & 0.003 & -103.67 & 0.44 \\
11h34m26.6s & -44d07m49.0s & 288.6375 & 16.5940 & 2.13 & 0.02 & 0.22 & 0.002 & -142.09 & 0.35 \\
11h34m24.6s & -44d06m43.0s & 288.6256 & 16.6096 & 2.12 & 0.01 & 0.37 & 0.002 & -136.83 & 0.20 \\
11h45m44.1s & -44d14m05.7s & 290.6960 & 17.0703 & 0.91 & 0.03 & 0.30 & 0.003 & -95.61 & 0.35 \\
11h46m30.3s & -43d51m44.4s & 290.7336 & 17.4661 & 0.16 & 0.06 & 0.02 & 0.002 & 176.67 & 3.95 \\
11h33m11.8s & -42d23m25.8s & 287.8465 & 18.1778 & 0.73 & 0.06 & 0.05 & 0.007 & 43.26 & 5.86 \\
11h23m10.5s & -41d41m20.7s & 285.7536 & 18.2197 & 1.31 & 0.02 & 0.05 & 0.002 & -54.49 & 1.41 \\
11h20m13.6s & -41d54m12.2s & 285.2944 & 17.8233 & 1.23 & 0.05 & 0.21 & 0.005 & -28.40 & 1.08 \\
11h20m13.6s & -41d54m12.3s & 285.2945 & 17.8233 & 1.41 & 0.03 & 0.26 & 0.002 & -29.54 & 0.33 \\
11h20m13.6s & -41d54m13.0s & 285.2946 & 17.8231 & 1.50 & 0.04 & 0.28 & 0.002 & -31.12 & 0.23 \\
11h15m46.6s & -39d14m21.2s & 283.4152 & 19.9799 & 1.17 & 0.06 & 0.21 & 0.003 & -60.55 & 0.64 \\
11h17m30.0s & -39d37m46.3s & 283.9001 & 19.7454 & 0.41 & 0.07 & 0.03 & 0.006 & 133.90 & 7.78 \\
11h18m14.5s & -39d23m12.8s & 283.9441 & 20.0240 & 0.67 & 0.08 & 0.14 & 0.002 & -58.58 & 0.54 \\
11h16m06.4s & -40d03m41.1s & 283.8111 & 19.2444 & 4.46 & 0.03 & 0.06 & 0.002 & -23.96 & 0.98 \\
11h26m44.2s & -38d28m45.7s & 285.2427 & 21.4655 & 13.56 & 0.02 & 0.68 & 0.001 & -45.10 & 0.08 \\
11h25m31.6s & -35d57m04.4s & 284.0378 & 23.7486 & 0.91 & 0.02 & 0.08 & 0.002 & -60.20 & 0.85 \\
11h29m46.2s & -39d06m51.8s & 286.0703 & 21.0685 & 0.43 & 0.01 & 0.01 & 0.001 & -295.37 & 3.39 \\
11h30m03.0s & -38d45m36.2s & 285.9990 & 21.4207 & 1.90 & 0.08 & 0.06 & 0.002 & -33.30 & 1.83 \\
11h26m44.2s & -38d28m44.6s & 285.2426 & 21.4658 & 15.10 & 0.03 & 0.52 & 0.004 & -46.36 & 0.28 \\
11h28m09.5s & -39d00m44.4s & 285.7180 & 21.0596 & 2.11 & 0.04 & 0.10 & 0.003 & 3.66 & 1.00 \\
11h28m14.9s & -39d08m31.4s & 285.7826 & 20.9434 & 2.85 & 0.06 & 0.07 & 0.003 & -28.07 & 1.66 \\
11h29m47.4s & -41d07m06.8s & 286.7800 & 19.1783 & 0.89 & 0.05 & 0.12 & 0.003 & -40.73 & 1.13 \\
11h30m10.1s & -40d56m46.7s & 286.7916 & 19.3644 & 1.52 & 0.02 & 0.07 & 0.002 & -36.88 & 1.25 \\
11h30m14.4s & -41d36m00.6s & 287.0310 & 18.7509 & 2.15 & 0.11 & 0.17 & 0.003 & -14.89 & 0.74 \\
11h33m17.9s & -43d44m56.1s & 288.3097 & 16.8931 & 1.40 & 0.04 & 0.10 & 0.002 & -117.34 & 0.64 \\
11h26m44.2s & -38d28m44.4s & 285.2427 & 21.4659 & 14.16 & 0.15 & 0.61 & 0.009 & -52.33 & 0.58 \\
11h26m44.2s & -38d28m46.0s & 285.2429 & 21.4654 & 5.80 & 0.06 & 0.41 & 0.005 & -46.57 & 0.53 \\
11h24m18.0s & -39d49m06.1s & 285.2681 & 20.0458 & 1.66 & 0.08 & 0.18 & 0.006 & -38.28 & 1.31 \\
11h23m22.4s & -39d42m40.4s & 285.0506 & 20.0831 & 0.67 & 0.07 & 0.08 & 0.003 & -22.89 & 1.42 \\
11h14m47.4s & -39d32m52.5s & 283.3545 & 19.6215 & 2.21 & 0.07 & 0.30 & 0.003 & -30.66 & 0.35 \\
11h12m50.9s & -40d09m15.5s & 283.2411 & 18.9178 & 1.50 & 0.04 & 0.26 & 0.002 & -36.30 & 0.32 \\
10h46m16.6s & -38d06m28.1s & 277.4074 & 18.4829 & 5.48 & 0.07 & 0.18 & 0.003 & -8.27 & 0.58 \\
10h46m10.8s & -38d05m52.1s & 277.3849 & 18.4824 & 2.72 & 0.02 & 0.29 & 0.002 & -13.91 & 0.29 \\
10h40m55.6s & -42d58m51.8s & 278.9739 & 13.7378 & 0.75 & 0.07 & 0.06 & 0.008 & -69.92 & 5.93 \\
10h42m17.4s & -43d07m06.7s & 279.2665 & 13.7404 & 3.15 & 0.05 & 0.13 & 0.005 & 329.60 & 1.52 \\
10h43m42.2s & -43d20m38.3s & 279.6110 & 13.6689 & 2.52 & 0.02 & 0.10 & 0.002 & -3.57 & 0.66 \\
10h54m04.3s & -44d57m29.8s & 282.0759 & 13.1081 & 2.76 & 0.04 & 0.26 & 0.004 & 28.85 & 0.62 \\
10h58m55.9s & -45d45m19.0s & 283.2219 & 12.7668 & 5.45 & 0.01 & 0.06 & 0.001 & -46.24 & 0.72 \\
10h55m22.0s & -43d35m41.7s & 281.6621 & 14.4313 & 2.51 & 0.02 & 0.13 & 0.002 & -11.00 & 0.57 \\
10h57m23.4s & -42d56m32.9s & 281.7047 & 15.1794 & 1.12 & 0.03 & 0.03 & 0.002 & 18.16 & 2.21 \\
10h56m43.2s & -42d40m07.5s & 281.4652 & 15.3708 & 0.73 & 0.12 & 0.03 & 0.009 & -147.07 & 12.64 \\
10h48m38.3s & -41d14m01.6s & 279.3956 & 15.9585 & 22.86 & 0.18 & 2.73 & 0.008 & -78.97 & 0.11 \\
10h37m24.1s & -39d35m41.7s & 276.6178 & 16.3315 & 4.54 & 0.05 & 0.16 & 0.003 & -54.46 & 0.70 \\
10h37m24.1s & -39d35m41.7s & 276.6178 & 16.3315 & 4.42 & 0.04 & 0.21 & 0.002 & -53.32 & 0.40 \\
10h39m39.9s & -38d08m31.6s & 276.2387 & 17.8057 & 2.74 & 0.07 & 0.30 & 0.005 & -2.62 & 0.63 \\
10h41m37.8s & -44d22m54.3s & 279.7947 & 12.5805 & 5.06 & 0.06 & 0.14 & 0.003 & -0.23 & 0.76 \\
10h41m37.8s & -44d22m54.3s & 279.7947 & 12.5805 & 5.21 & 0.01 & 0.24 & 0.001 & -13.13 & 0.15 \\
10h38m01.0s & -42d08m51.5s & 278.0665 & 14.1922 & 1.94 & 0.03 & 0.05 & 0.003 & -59.89 & 2.47 \\
11h04m31.0s & -42d34m45.3s & 282.7705 & 16.0680 & 0.52 & 0.02 & 0.03 & 0.002 & 33.88 & 2.54 \\
11h04m31.0s & -42d34m46.1s & 282.7706 & 16.0678 & 4.93 & 0.16 & 0.47 & 0.002 & 29.77 & 0.18 \\
11h07m05.8s & -48d08m26.9s & 285.5291 & 11.1739 & 5.88 & 0.03 & 0.08 & 0.003 & -60.63 & 1.44 \\
11h33m44.8s & -47d20m43.5s & 289.5336 & 13.4951 & 9.21 & 0.05 & 0.13 & 0.002 & -28.78 & 0.62 \\
11h05m50.9s & -48d57m09.1s & 285.6674 & 10.3474 & 0.50 & 0.06 & 0.06 & 0.005 & -88.24 & 3.46 \\
11h04m30.2s & -48d56m16.2s & 285.4560 & 10.2714 & 0.68 & 0.02 & 0.04 & 0.002 & -110.45 & 1.61 \\
10h47m02.0s & -47d36m56.0s & 282.2128 & 10.1891 & 1.34 & 0.03 & 0.30 & 0.003 & -23.44 & 0.39 \\
10h30m41.2s & -44d53m47.6s & 278.3453 & 11.1513 & 2.21 & 0.03 & 0.03 & 0.001 & -43.77 & 2.08 \\
10h08m05.4s & -41d21m22.2s & 272.8396 & 11.7788 & 2.73 & 0.04 & 0.04 & 0.002 & -1.83 & 2.10 \\
10h00m44.9s & -41d59m49.7s & 272.1207 & 10.4400 & 1.67 & 0.02 & 0.06 & 0.003 & 348.56 & 1.87 \\
10h11m45.7s & -41d28m53.0s & 273.4836 & 12.0834 & 1.06 & 0.03 & 0.18 & 0.003 & -171.15 & 0.74 \\
10h11m46.4s & -41d27m34.0s & 273.4723 & 12.1025 & 1.66 & 0.03 & 0.09 & 0.003 & -139.11 & 1.15 \\
10h17m17.2s & -40d47m56.4s & 273.9509 & 13.2407 & 4.28 & 0.03 & 0.03 & 0.002 & -141.09 & 2.27 \\
10h26m48.0s & -41d43m24.6s & 276.0069 & 13.4639 & 8.69 & 0.05 & 0.30 & 0.004 & -174.52 & 0.57 \\
10h26m49.3s & -41d43m02.6s & 276.0068 & 13.4712 & 8.61 & 0.07 & 0.61 & 0.004 & -186.46 & 0.26 \\
10h38m17.3s & -45d09m39.4s & 279.6610 & 11.6101 & 1.51 & 0.04 & 0.02 & 0.002 & 51.83 & 4.63 \\
11h07m23.0s & -42d21m55.8s & 283.1808 & 16.4806 & 2.14 & 0.10 & 0.16 & 0.011 & 17.68 & 3.02 \\
11h08m17.1s & -42d07m48.2s & 283.2400 & 16.7634 & 1.97 & 0.04 & 0.13 & 0.003 & -5.94 & 0.88 \\
11h03m28.5s & -41d04m10.5s & 281.9248 & 17.3567 & 3.38 & 0.11 & 0.23 & 0.003 & 20.96 & 0.55 \\
11h11m19.8s & -40d30m41.3s & 283.1095 & 18.4755 & 18.08 & 0.02 & 1.04 & 0.003 & -25.12 & 0.12 \\
11h38m01.5s & -39d22m53.2s & 287.7869 & 21.3216 & 6.05 & 0.01 & 0.69 & 0.002 & 6.71 & 0.11 \\
11h41m50.5s & -35d04m11.5s & 287.1772 & 25.6637 & 4.79 & 0.08 & 0.17 & 0.007 & 5.27 & 1.72 \\
11h41m50.2s & -35d04m05.8s & 287.1755 & 25.6649 & 9.00 & 0.06 & 0.16 & 0.004 & -17.21 & 0.94 \\
11h44m07.4s & -39d22m52.7s & 288.9993 & 21.6623 & 0.72 & 0.06 & 0.13 & 0.003 & 83.99 & 1.17 \\
11h44m30.9s & -34d57m57.9s & 287.7254 & 25.9218 & 1.01 & 0.04 & 0.11 & 0.005 & 7.35 & 1.78 \\
11h45m47.6s & -31d59m01.3s & 287.0527 & 28.8559 & 7.11 & 0.11 & 0.66 & 0.016 & -35.76 & 0.95 \\
11h45m01.6s & -39d09m17.1s & 289.1130 & 21.9283 & 4.89 & 0.06 & 0.18 & 0.003 & 55.67 & 0.68 \\
11h47m44.4s & -38d32m43.3s & 289.4856 & 22.6575 & 2.38 & 0.04 & 0.10 & 0.003 & -8.75 & 1.35 \\
11h47m29.8s & -36d03m06.0s & 288.7049 & 25.0473 & 1.56 & 0.04 & 0.04 & 0.003 & -15.02 & 2.80 \\
11h46m36.5s & -37d57m21.1s & 289.0819 & 23.1661 & 0.42 & 0.02 & 0.03 & 0.006 & 3.60 & 8.10 \\
11h47m01.5s & -38d12m11.7s & 289.2400 & 22.9499 & 33.29 & 0.19 & 1.16 & 0.009 & -4.55 & 0.31 \\
11h47m44.3s & -38d32m44.3s & 289.4854 & 22.6572 & 2.30 & 0.01 & 0.13 & 0.002 & 8.52 & 0.56 \\
11h47m53.9s & -38d24m11.4s & 289.4773 & 22.8029 & 3.85 & 0.04 & 0.17 & 0.003 & 2.68 & 0.74 \\
11h47m54.0s & -38d24m11.9s & 289.4774 & 22.8028 & 3.23 & 0.04 & 0.42 & 0.003 & 3.88 & 0.33 \\
11h49m10.6s & -32d59m13.7s & 288.1504 & 28.0912 & 1.14 & 0.09 & 0.04 & 0.008 & -377.82 & 7.88 \\
11h49m08.5s & -35d25m32.0s & 288.8754 & 25.7400 & 0.44 & 0.05 & 0.04 & 0.008 & 126.91 & 7.53 \\
11h50m35.6s & -38d30m29.1s & 290.0595 & 22.8380 & 0.49 & 0.04 & 0.07 & 0.009 & 24.74 & 4.77 \\
11h50m31.1s & -37d59m17.9s & 289.9006 & 23.3371 & 1.06 & 0.02 & 0.06 & 0.002 & 48.33 & 1.22 \\
11h51m25.7s & -37d57m19.7s & 290.0807 & 23.4142 & 2.48 & 0.03 & 0.05 & 0.002 & 59.80 & 1.75 \\
11h52m19.4s & -36d10m06.6s & 289.7782 & 25.1888 & 0.45 & 0.05 & 0.05 & 0.004 & 16.83 & 3.13 \\
11h54m01.6s & -35d32m23.2s & 289.9759 & 25.8834 & 2.40 & 0.11 & 0.56 & 0.013 & 34.94 & 0.97 \\
11h53m13.8s & -37d14m15.7s & 290.2639 & 24.1981 & 3.58 & 0.04 & 0.45 & 0.006 & -34.54 & 0.53 \\
11h55m55.8s & -36d56m39.7s & 290.7603 & 24.6120 & 2.28 & 0.05 & 0.05 & 0.004 & 19.40 & 3.30 \\
11h56m16.9s & -36d41m20.0s & 290.7699 & 24.8768 & 0.38 & 0.05 & 0.05 & 0.007 & 65.09 & 6.35 \\
11h24m50.6s & -33d54m58.0s & 283.0878 & 25.5968 & 0.92 & 0.06 & 0.06 & 0.005 & 218.09 & 3.63 \\
12h01m18.7s & -35d13m54.8s & 291.5071 & 26.5254 & 2.15 & 0.07 & 0.14 & 0.012 & -41.06 & 3.99 \\
12h01m18.7s & -35d13m55.3s & 291.5072 & 26.5252 & 1.61 & 0.09 & 0.16 & 0.007 & -39.53 & 1.86 \\
11h28m13.7s & -35d59m45.6s & 284.6145 & 23.8985 & 0.29 & 0.06 & 0.04 & 0.005 & -191.43 & 4.39 \\
12h03m37.5s & -34d23m18.1s & 291.8273 & 27.4503 & 0.12 & 0.02 & 0.04 & 0.003 & 22.52 & 2.93 \\
11h30m41.3s & -34d10m34.4s & 284.4470 & 25.7804 & 3.69 & 0.12 & 0.08 & 0.009 & -325.25 & 4.17 \\
11h31m28.2s & -34d02m43.1s & 284.5667 & 25.9585 & 0.71 & 0.03 & 0.07 & 0.002 & -26.02 & 1.11 \\
11h32m24.0s & -35d07m29.9s & 285.1672 & 25.0054 & 0.66 & 0.09 & 0.05 & 0.007 & 126.81 & 5.37 \\
12h09m17.1s & -34d45m32.6s & 293.1956 & 27.3145 & 1.03 & 0.08 & 0.11 & 0.009 & -146.09 & 3.36 \\
11h34m35.1s & -32d49m16.5s & 284.7943 & 27.3293 & 10.97 & 0.05 & 0.49 & 0.007 & -78.75 & 0.54 \\
12h13m43.0s & -36d33m28.0s & 294.5333 & 25.6981 & 2.34 & 0.07 & 0.23 & 0.005 & -136.15 & 0.88 \\
11h35m15.3s & -35d15m32.4s & 285.8240 & 25.0691 & 2.14 & 0.09 & 0.13 & 0.007 & -23.73 & 2.30 \\
11h36m51.0s & -37d15m49.5s & 286.8512 & 23.2679 & 0.68 & 0.08 & 0.06 & 0.003 & -30.99 & 2.26 \\
12h07m42.6s & -45d30m22.8s & 294.9563 & 16.6872 & 1.44 & 0.11 & 0.09 & 0.004 & -124.70 & 1.81 \\
12h05m22.2s & -43d55m19.9s & 294.2262 & 18.1694 & 2.42 & 0.04 & 0.35 & 0.003 & -96.63 & 0.35 \\
12h02m24.9s & -43d50m30.7s & 293.6602 & 18.1456 & 6.57 & 0.02 & 0.59 & 0.002 & -67.54 & 0.10 \\
12h02m24.9s & -43d50m31.0s & 293.6601 & 18.1455 & 4.45 & 0.08 & 0.46 & 0.006 & -66.25 & 0.53 \\
12h14m18.3s & -43d15m03.4s & 295.7889 & 19.1052 & 1.51 & 0.08 & 0.28 & 0.002 & -107.48 & 0.28 \\
11h49m54.7s & -39d33m40.1s & 290.2101 & 21.7859 & 0.60 & 0.03 & 0.13 & 0.003 & 25.12 & 0.76 \\
11h49m15.0s & -39d40m51.2s & 290.1103 & 21.6375 & 0.36 & 0.07 & 0.05 & 0.003 & 28.22 & 2.42 \\
11h49m54.7s & -39d33m39.8s & 290.2101 & 21.7860 & 0.45 & 0.02 & 0.02 & 0.001 & 2.81 & 2.01 \\
11h47m06.6s & -41d11m16.7s & 290.1081 & 20.0764 & 17.87 & 0.16 & 1.03 & 0.004 & 37.76 & 0.15 \\
11h47m06.7s & -41d11m15.9s & 290.1083 & 20.0767 & 16.32 & 0.03 & 0.97 & 0.001 & 37.12 & 0.06 \\
11h46m29.6s & -42d34m17.2s & 290.3760 & 18.7114 & 2.25 & 0.04 & 0.20 & 0.004 & 389.41 & 0.82 \\
12h01m38.7s & -42d48m40.3s & 293.2969 & 19.1271 & 1.11 & 0.05 & 0.04 & 0.003 & -57.05 & 3.73 \\
12h06m52.7s & -42d52m54.0s & 294.3079 & 19.2427 & 8.04 & 0.02 & 0.05 & 0.002 & -75.08 & 1.88 \\
12h14m16.8s & -42d00m33.2s & 295.5831 & 20.3317 & 2.92 & 0.03 & 0.68 & 0.004 & -130.06 & 0.22 \\
12h25m58.7s & -43d14m43.7s & 298.0210 & 19.3834 & 0.61 & 0.03 & 0.05 & 0.002 & -29.37 & 1.70 \\
12h27m55.6s & -44d20m57.8s & 298.5077 & 18.3203 & 6.16 & 0.02 & 0.75 & 0.001 & -44.12 & 0.06 \\
12h27m55.6s & -44d19m55.8s & 298.5059 & 18.3374 & 6.76 & 0.01 & 0.29 & 0.001 & -47.82 & 0.15 \\
12h23m24.8s & -49d29m47.7s & 298.2660 & 13.1211 & 1.61 & 0.06 & 0.10 & 0.001 & 17.19 & 0.49 \\
12h31m26.2s & -48d53m23.8s & 299.5481 & 13.8503 & 1.46 & 0.03 & 0.04 & 0.002 & -39.02 & 1.98 \\
12h35m18.9s & -46d00m40.7s & 300.0093 & 16.7689 & 1.11 & 0.04 & 0.11 & 0.003 & 0.57 & 0.94 \\
12h15m59.6s & -42d30m28.2s & 295.9962 & 19.8861 & 1.46 & 0.02 & 0.12 & 0.001 & -74.21 & 0.50 \\
12h11m14.5s & -39d33m27.4s & 294.5492 & 22.6590 & 0.36 & 0.07 & 0.04 & 0.006 & -158.61 & 6.44 \\
12h49m35.4s & -48d16m54.3s & 302.6143 & 14.5889 & 1.29 & 0.04 & 0.01 & 0.001 & 71.07 & 2.98 \\
12h50m14.1s & -48d08m42.0s & 302.7244 & 14.7262 & 0.48 & 0.03 & 0.03 & 0.002 & 30.70 & 2.33 \\
12h50m14.1s & -48d08m42.8s & 302.7245 & 14.7261 & 0.72 & 0.04 & 0.05 & 0.005 & 39.44 & 4.21 \\
12h48m12.7s & -47d47m15.8s & 302.3706 & 15.0805 & 0.21 & 0.01 & 0.01 & 0.003 & -200.26 & 10.11 \\
12h19m01.6s & -44d09m42.1s & 296.8173 & 18.3249 & 0.47 & 0.05 & 0.04 & 0.002 & -65.48 & 2.40 \\
12h18m32.2s & -38d55m51.7s & 295.9617 & 23.4950 & 1.09 & 0.03 & 0.14 & 0.003 & -88.42 & 0.94 \\
12h16m29.8s & -38d54m10.2s & 295.5293 & 23.4662 & 0.84 & 0.06 & 0.04 & 0.002 & -94.46 & 2.17 \\
12h15m19.1s & -39d04m28.1s & 295.3106 & 23.2622 & 1.09 & 0.04 & 0.06 & 0.002 & -140.68 & 1.49 \\
12h15m59.6s & -42d30m28.4s & 295.9962 & 19.8860 & 1.04 & 0.03 & 0.06 & 0.002 & -79.62 & 1.27 \\
11h50m55.4s & -46d32m49.8s & 292.2194 & 15.0620 & 2.06 & 0.05 & 0.06 & 0.002 & -81.09 & 1.70 \\
12h04m59.4s & -38d50m32.9s & 293.1153 & 23.1423 & 2.46 & 0.04 & 0.16 & 0.002 & -4.41 & 0.49 \\
11h58m48.8s & -40d30m06.9s & 292.2286 & 21.2736 & 20.63 & 0.06 & 0.16 & 0.003 & -27.60 & 0.81 \\
11h54m17.8s & -40d47m28.8s & 291.4046 & 20.7979 & 8.28 & 0.11 & 0.58 & 0.010 & -45.71 & 0.73 \\
11h47m06.7s & -41d11m16.5s & 290.1084 & 20.0766 & 21.95 & 0.05 & 1.18 & 0.006 & 40.19 & 0.21 \\
11h51m21.2s & -40d50m17.8s & 290.8389 & 20.6183 & 1.12 & 0.03 & 0.06 & 0.002 & -45.11 & 1.22 \\
11h50m07.5s & -41d14m59.1s & 290.7087 & 20.1622 & 0.88 & 0.07 & 0.05 & 0.002 & -14.68 & 1.95 \\
11h44m08.8s & -40d15m17.1s & 289.2629 & 20.8242 & 2.52 & 0.03 & 0.10 & 0.003 & 138.70 & 1.21 \\
11h44m48.2s & -40d15m07.1s & 289.3910 & 20.8612 & 2.21 & 0.07 & 0.16 & 0.004 & 129.10 & 1.10 \\
11h44m08.8s & -40d15m16.8s & 289.2629 & 20.8242 & 2.93 & 0.13 & 0.11 & 0.004 & 136.05 & 1.55 \\
11h44m48.2s & -40d15m06.8s & 289.3910 & 20.8613 & 2.06 & 0.03 & 0.15 & 0.003 & 117.67 & 0.86 \\
11h45m57.3s & -41d34m39.8s & 289.9955 & 19.6432 & 2.11 & 0.10 & 0.16 & 0.003 & -9.56 & 0.67 \\
11h47m06.7s & -41d11m15.8s & 290.1083 & 20.0767 & 18.02 & 0.01 & 1.00 & 0.002 & 34.06 & 0.10 \\
11h49m09.8s & -41d06m59.3s & 290.4861 & 20.2454 & 1.17 & 0.05 & 0.12 & 0.004 & -31.87 & 1.21 \\
11h51m57.0s & -40d26m32.5s & 290.8521 & 21.0299 & 3.71 & 0.02 & 0.07 & 0.003 & 101.34 & 1.81 \\
12h40m13.3s & -43d05m08.4s & 300.7566 & 19.7387 & 1.52 & 0.06 & 0.07 & 0.004 & -34.06 & 2.36 \\
12h44m33.3s & -50d10m10.7s & 301.8022 & 12.6870 & 0.78 & 0.02 & 0.09 & 0.001 & 14.77 & 0.47 \\
12h44m51.9s & -50d09m39.5s & 301.8530 & 12.6970 & 3.22 & 0.02 & 0.22 & 0.001 & 3.99 & 0.22 \\
12h37m59.5s & -50d57m12.6s & 300.7689 & 11.8618 & 1.74 & 0.04 & 0.06 & 0.001 & 35.12 & 0.75 \\
12h28m26.5s & -51d49m40.0s & 299.3174 & 10.8825 & 2.70 & 0.04 & 0.40 & 0.003 & 51.52 & 0.34 \\
11h57m23.0s & -45d05m57.5s & 293.0139 & 16.7289 & 2.67 & 0.05 & 0.13 & 0.003 & -113.87 & 1.09 \\
11h49m46.2s & -43d45m37.8s & 291.3039 & 17.7145 & 4.69 & 0.03 & 0.14 & 0.003 & -57.36 & 1.03 \\
11h50m11.6s & -43d30m13.2s & 291.3149 & 17.9823 & 1.38 & 0.03 & 0.16 & 0.003 & -95.25 & 0.64 \\
11h49m46.2s & -43d45m38.1s & 291.3039 & 17.7144 & 6.88 & 0.13 & 0.28 & 0.009 & -56.66 & 1.25 \\
11h50m11.6s & -43d30m13.6s & 291.3149 & 17.9822 & 1.05 & 0.03 & 0.06 & 0.003 & -69.51 & 1.97 \\
11h50m06.6s & -43d29m56.6s & 291.2984 & 17.9830 & 0.77 & 0.03 & 0.09 & 0.002 & -62.38 & 1.11 \\
12h05m19.3s & -43d55m44.1s & 294.2187 & 18.1612 & 2.65 & 0.04 & 0.85 & 0.003 & -95.86 & 0.14 \\
12h23m30.5s & -43d59m04.6s & 297.6366 & 18.5998 & 0.90 & 0.01 & 0.13 & 0.002 & -74.22 & 0.53 \\
12h42m01.4s & -45d35m09.8s & 301.2075 & 17.2542 & 0.87 & 0.02 & 0.02 & 0.001 & -52.19 & 3.30 \\
12h44m04.6s & -44d51m41.2s & 301.5605 & 17.9907 & 0.86 & 0.02 & 0.04 & 0.002 & -59.70 & 2.23 \\
12h37m49.8s & -51d43m46.0s & 300.7856 & 11.0856 & 3.94 & 0.02 & 0.08 & 0.001 & -42.52 & 0.40 \\
12h13m14.3s & -49d59m35.0s & 296.6620 & 12.4131 & 4.10 & 0.07 & 0.21 & 0.005 & -4.15 & 1.01 \\
12h06m30.6s & -50d34m42.9s & 295.6764 & 11.6592 & 4.93 & 0.03 & 0.04 & 0.001 & -88.92 & 1.47 \\
12h06m10.9s & -50d28m22.4s & 295.6048 & 11.7539 & 0.70 & 0.03 & 0.02 & 0.002 & -100.04 & 3.37 \\
11h55m04.8s & -50d51m30.9s & 293.9249 & 11.0249 & 1.11 & 0.03 & 0.07 & 0.001 & -73.91 & 0.74 \\
11h43m53.9s & -48d04m33.1s & 291.4303 & 13.2818 & 1.69 & 0.02 & 0.09 & 0.001 & 10.49 & 0.69 \\
11h56m32.5s & -44d38m57.7s & 292.7589 & 17.1353 & 1.58 & 0.03 & 0.04 & 0.002 & -47.23 & 2.09 \\
11h57m56.5s & -38d56m45.8s & 291.6824 & 22.7543 & 1.34 & 0.03 & 0.12 & 0.002 & -23.88 & 0.67 \\
11h56m41.9s & -39d14m39.4s & 291.5007 & 22.4093 & 1.00 & 0.03 & 0.59 & 0.002 & 12.28 & 0.16 \\
12h10m35.7s & -40d28m18.3s & 294.5843 & 21.7371 & 0.95 & 0.05 & 0.05 & 0.009 & -187.31 & 7.82 \\
12h10m39.1s & -41d10m19.8s & 294.7231 & 21.0486 & 1.39 & 0.03 & 0.18 & 0.002 & -83.10 & 0.52 \\
12h32m33.9s & -45d52m38.8s & 299.5012 & 16.8685 & 9.47 & 0.03 & 0.07 & 0.002 & 25.24 & 1.18 \\
12h42m53.3s & -40d28m15.6s & 301.1736 & 22.3716 & 1.16 & 0.02 & 0.07 & 0.002 & -22.16 & 1.38 \\
13h11m21.5s & -43d00m24.4s & 306.7985 & 19.7154 & 2.37 & 0.04 & 0.03 & 0.003 & -28.88 & 4.18 \\
13h02m14.7s & -47d22m04.7s & 304.8302 & 15.4639 & 1.96 & 0.04 & 0.42 & 0.002 & -0.51 & 0.20 \\
13h05m29.7s & -48d18m56.6s & 305.3450 & 14.4902 & 1.92 & 0.02 & 0.05 & 0.001 & 134.11 & 0.62 \\
12h56m02.6s & -48d18m13.0s & 303.7232 & 14.5611 & 2.94 & 0.04 & 0.25 & 0.002 & 158.91 & 0.25 \\
12h54m59.6s & -48d12m25.8s & 303.5442 & 14.6603 & 0.24 & 0.02 & 0.02 & 0.002 & 421.99 & 4.08 \\
13h02m28.1s & -44d47m36.6s & 304.9895 & 18.0342 & 14.77 & 0.06 & 0.61 & 0.003 & -47.63 & 0.20 \\
12h59m57.3s & -44d06m30.4s & 304.5461 & 18.7366 & 0.83 & 0.06 & 0.03 & 0.002 & -41.20 & 2.83 \\
12h51m36.7s & -41d33m42.9s & 302.9667 & 21.3098 & 3.37 & 0.02 & 0.44 & 0.003 & -54.80 & 0.25 \\
\hline
\end{longtable}

\bsp	
\label{lastpage}
\end{document}